\documentclass[twocolumn,showpacs,floatfix,pre]{revtex4}
\usepackage{graphicx}
\usepackage{dcolumn}
\usepackage{bm}
\def\tr{t_{\rm r}}
\def\om{\omega_F}

\def\sgn{{\rm sgn}}

\begin{document}
\title{Scaling and crossovers in activated escape near a bifurcation point}

\author{D. Ryvkine, M.I. Dykman,
and B. Golding}
 \affiliation{Department of Physics and Astronomy, Michigan State
 University}
\date{\today}

\begin{abstract}
Near a bifurcation point a system experiences critical slowing down.
This leads to scaling behavior of fluctuations. We find that a
periodically driven system may display three scaling regimes and
scaling crossovers near a saddle-node bifurcation where a metastable
state disappears.  The rate of activated escape $W$ scales with the
driving field amplitude $A$ as $\ln W \propto (A_c-A)^{\xi}$, where
$A_c$ is the bifurcational value of $A$. With increasing field
frequency the critical exponent $\xi$ changes from $\xi = 3/2$ for
stationary systems to a dynamical value $\xi=2$ and then again to
$\xi=3/2$. The analytical results are in agreement with the results of
asymptotic calculations in the scaling region. Numerical
calculations and simulations for a model system support the theory.
\end{abstract}


\maketitle

\section{Introduction}

Thermally activated transitions are at root of many physical
phenomena: diffusion in solids, protein folding, and nucleation are
examples. It is important to understand how transitions occur,
particularly in systems away from thermal equilibrium. Full
understanding would include a description of the underlying dynamics
and the transition probabilities. Owing to their exponential
sensitivity, these probabilities provide an important means of
characterizing a system. However, in many cases activation barriers
are high which leads to very low transition rates and impedes precise
experimental studies.

The barrier for escape from a metastable state is reduced when the
system is close to a bifurcation (critical, or spinodal) point where
the state disappears. For systems that display hysteresis such a
bifurcation point corresponds to the switching point on the hysteresis
loop. The idea of bringing the system close to the bifurcation point
\cite{Kurkijarvi72} has been used in studying activated switching in
Josephson junctions \cite{Fulton74,Martinis87,Sharifi88,Han89}, where
it has become a standard technique for determining the critical
current. This idea is also used in studies of activated magnetization
reversals in nanomagnets \cite{Wernsdorfer97,Koch00,Ralph02}.

Experiments on nanomagnets and Josephson junctions are often performed
by ramping the control parameter (magnetic field or current) and
measuring time distribution of escape events \cite{Kurkijarvi72}. In
interpreting the data it is usually assumed that, for sufficiently
slow ramp rates, the system remains quasistationary. In this
approximation the barrier height, i.e., the activation energy of a
transition $R$, usually scales with the control parameter $\eta$,
measured from its critical (bifurcational) value $\eta_c=0$, as
$\eta^{3/2}$ \cite{3/2,Ott03}.

Scaling of $R$ near a bifurcation point is related to slowing
down of one of the motions \cite{Guckenheimer}, i.e., the onset
of a ``soft mode''. The relaxation time of the system $\tr$
diverges as the control parameter $\eta\to 0$. Therefore if
$\eta$ depends on time, even where this dependence is slow the
assumption of quasistationary may become inapplicable for small
$\eta$.

In this paper a theory of activated transitions is developed for
periodically modulated systems. In such systems the notion of a stable
state is well-defined irrespective of the modulation rate, and the
applicability of the quasistationary approximation can be carefully
studied. It turns out that, unexpectedly, near a critical point this
approximation breaks down even where the relaxation time $\tr$ is still
much smaller than the driving period $\tau_F=2\pi/\omega_F$.

We show that an interplay between the critical slowing down and the
slowness of time-dependent modulation leads to a rich scaling behavior
of the transition rate and to crossovers between different scaling
regions. This behavior near a bifurcation point is system-independent
and has no counterparts in stationary systems.  We find three regions
in which the activation energy scales as $R\propto \eta^{\xi}$. As the
parameters change, for example with the increase of the modulation
frequency $\omega_F$, the critical exponent $\xi$ varies from 3/2 to 2
and then again to 3/2. Our numerical calculations and Monte Carlo
simulations for a model system agree with the general results.

Activated transitions in periodically driven systems were investigated
earlier in various contexts
\cite{Larkin,Devoret87,Gitterman,prefactor,Hanggi-00,M&S-01,LS_chaos,Soskin_RPP,Fistul_03},
stochastic resonance and diffusion in modulated ratchets being recent
examples \cite{SR-review,Wiesenfeld98,ratchets}.  In this paper we
study the previously unexplored region of driving amplitudes close to
critical and reveal the universality that emerges.

\begin{figure}[ht]
\includegraphics[width=3in]{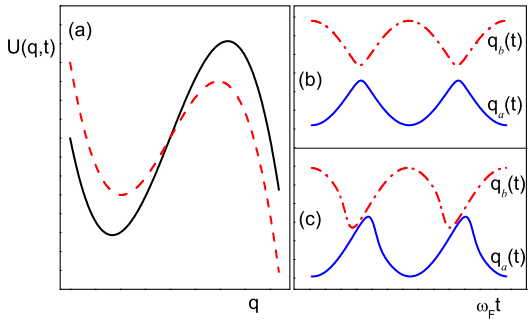}
\caption{(a) An oscillating potential barrier. In
the limit of slow modulation, the stable and unstable periodic
states $q_a$ and $q_b$ are the instantaneous positions of the
potential minimum and barrier top, respectively. (b) For slow
modulation, when the driving amplitude $A$ is close to $A_c^{\rm
ad}$, the states $q_{a,b}(t)$ come close to each other once per
period. (c) As $A$ further approaches $A_c$, the states
$q_{a,b}(t)$ become skewed compared to the adiabatic picture, to
avoid crossing.  In the critical range, the form of $q_{a,b}(t)$
is model-independent.}
\label{fig:barrier}
\end{figure}

A qualitative picture of motion near a bifurcation point can be
obtained if one thinks of the system as a particle in a potential
$U(q,t)$ that oscillates in time with period $\tau_F$, see
Fig.~\ref{fig:barrier}. Such particle has periodic stable and unstable
states, $q_a(t)$ and $q_b(t)$. In the adiabatic limit $\omega_F \to 0$
they lie at the minimum and local maximum of the potential in
Fig.~\ref{fig:barrier}a. As the modulation amplitude $A$ increases,
the states become close to each other for a portion of the period
$\tau_F$, see Fig.~\ref{fig:barrier}b. The barrier height reaches its
minimum during this time, and this is when the system is most likely
to escape from the potential well.

As $A$ further increases, the states $q_{a,b}(t)$ become distorted to
avoid crossing, see Fig.~\ref{fig:barrier}c, and the adiabatic
approximation becomes inapplicable. The parameter range where it
happens can be estimated by noticing that the adiabatic relaxation
time $\tr$ (i) is a function of the instantaneous modulation phase
$\phi=\omega_Ft$, and (ii)  sharply increases near the
bifurcation point. As a consequence, $\tr$ sharply increases when
$\phi$ approaches the value where $q_{a,b}$ are at their closest,
because this corresponds to approaching the bifurcation point. The
quasistationary (adiabatic) approximation requires that
$|\partial\tr/\partial t|\ll 1$. It is this condition that limits the
range of adiabaticity, rather than a much less restrictive condition
$\tr\omega_F\ll 1$.

Even for $A > A_c^{\rm ad}$, where the adiabatic barrier in
Fig.~\ref{fig:barrier} disappears for a portion of the period, the
states $q_{a,b}$ may still coexist. A sufficiently large fluctuation
is then required to move the system away from the stable periodic
state. It is in this region that the new scaling of the activation
energy $R$ emerges. The control parameter is now $\eta \propto A_c-A$,
where $A_c$ is the ``true'' bifurcation value of the modulation
amplitude.

In the limit $\omega_F\tr \gg 1$ the behavior near a bifurcation point
is in some sense simpler. In this case $q_a(t)$ and $q_b(t)$ come
close to each other everywhere on the cycle, not just for a part of
the period. The motion of the system in the vicinity of $q_{a,b}(t)$
is oscillations about the state $q_a(t)$ with a slowly varying
amplitude. The amplitude change can be described by averaging the
complete dynamics over the period. It is then mapped onto motion in an
effectively stationary potential. Not surprisingly, the scaling of the
escape activation energy $R$ with the distance to the bifurcation
point is the same as for stationary systems.

In Sec.~II and Appendix~A we provide a general formulation of the
problem of activated escape in periodically modulated systems driven
by Gaussian noise. In Sec.~III~A and Appendix~B we discuss the
dynamics near a bifurcation point in the adiabatic limit $\omega_F\tr
\to 0$.  In Sec.~III~B we consider the strongly nonadiabatic dynamics
that emerges where still $\omega_F\tr \ll 1$. In Sec.~III~C and
Appendix~C the dynamics near a bifurcation point is described in the
parameter range where the field becomes effectively fast-oscillating,
i.e. $\omega_F\tr \gg 1$, even though the relaxation time in the
absence of modulation $\tr^{(0)}$ may be $\lesssim 1/\omega_F$. The
connection between the nonadiabatic local theory of Sec.~III~B and the
theory of Sec.~III~C is discussed in Sec.~III~D. In Sec.~IV the
activation energy is explicitly evaluated in the three regions
discussed in Sec.~III, and the scaling laws for the activation energy
$R\propto (A_c-A)^{\xi}$ in these regions are obtained. The scaling
crossovers are discussed. We also find nonadiabatic corrections to the
escape rate in the adiabatic region. In Sec.~V we consider a
periodically modulated Brownian particle. Numerical results for the
activation energy are compared to the results of Monte Carlo
simulations and to the predictions of Sec.~IV. Sec.~VI contains
concluding remarks.

\section{Activated escape: general formulation}

We will adopt a phenomenological approach in which a multidimensional
system with dynamical variables ${\bf q}(t)$ is described by the
Langevin equation
\begin{equation}
\label{eom}
\dot{\bf q} = {\bf K}({\bf q}; A,t) + {\bf f}(t), \quad {\bf K}({\bf
q};A,t+\tau_F) = {\bf K}({\bf q}; A,t).
\end{equation}
The function ${\bf K}$ is periodic in time, with the modulation period
$\tau_F=2\pi/\omega_F$; $A$
is a control parameter that characterizes the modulation strength. For
example, in the case of an overdamped particle in a potential $U_0({\bf
q})$ modulated by an additive periodic force ${\bf F}(t)$, the vector
${\bf K}$ becomes
\begin{equation}
\label{additive_force}
{\bf K}({\bf q}; A,t) = - \bm{\nabla}U_0({\bf q}) + {\bf F}(t).
\end{equation}
(here and below, $\bm{\nabla}\equiv \partial/\partial {\bf q}$). In
this case $A=\max |{\bf F}|$ is the modulation amplitude (note that
the force ${\bf F}(t)={\bf F}(t+\tau_F)$ does not have to be
sinusoidal).

The function ${\bf f}(t)$ in Eq.~(\ref{eom}) is zero-mean
Gaussian noise with correlation matrix
\begin{equation}
\label{corr_matrix}
\varphi_{ij}(t-t')=\langle f_i(t)f_j(t')\rangle.
\end{equation}
The characteristic noise intensity $D$ can be defined as the maximal
value of  the power spectrum,
\begin{equation}
\label{noise_intensity}
D=\max\Phi_{nn}(\omega), \quad
\Phi_{nm}(\omega)=\int dt e^{i\omega t}\varphi_{nm}(t).
\end{equation}
For thermal fluctuations $D=2k_BT$.  The noise intensity $D$ is the
smallest parameter of the theory.  Smallness of $D$ leads to the rate
of noise-induced escape $W$ being much smaller than
$\tr^{-1}$ and $\omega_F$.

In the absence of noise, Eq.~(\ref{eom}) may have different
periodic solutions ${\bf q}_{\rm per}$, which can be stable
(attractors), unstable (repellers), or hyperbolic (saddles). We
are interested in the parameter range where one of the stable
periodic solutions ${\bf q}_a(t)= {\bf q}_a(t+\tau_F)$ comes
close to a saddle-type periodic solution ${\bf q}_b(t)$ with the
same period (period 1, for
 concreteness). For slow modulation, these states are
sketched in Fig.~\ref{fig:barrier}. They
merge together at the saddle-node bifurcation point $A=A_c$. In
what follows we will assume that $A$ is close to the critical value
$A_c$.

Escape from a metastable state ${\bf q}_a(t)$ occurs as a result of a
large fluctuation. The fluctuational force ${\bf f}(t)$ has to
overcome the restoring force ${\bf K}$ and drive the system away from
the basin of attraction
to ${\bf q}_a(t)$
[e.g., away from the potential well in
Fig.~\ref{fig:barrier}].  We will assume that the required force ${\bf
f}(t)$ is much larger than the typical noise amplitude $\propto
D^{1/2}$.

The motion of the system during escape is random. However, different
trajectories have exponentially different probabilities. The system is
most likely to move along a particular trajectory called the optimal
path ${\bf q}_{\rm opt}(t)$ \cite{Onsager}. It is determined by the most
probable noise realization ${\bf f}_{\rm opt}(t)$.
In the case of a periodically modulated 1D system driven by stationary
Gaussian noise, a way to find the optimal paths was discussed earlier
\cite{LS_chaos}. We now briefly outline a generalization of the
formulation to multidimensional systems, following the arguments in
Ref.~\onlinecite{Dykman-90} (more details are provided in Appendix A).

For a stationary Gaussian noise, the probability density of
realizations of ${\bf f}(t)$ is given by the functional
(cf. Ref. \onlinecite{Feynman_book})
\begin{equation}
\label{P_functional}
\mathcal{P}[{\bf f}(t)]= \exp(-\mathcal{R}_0[{\bf f}(t)]/D),
\end{equation}
where ${\cal R}_0$ is quadratic in ${\bf f}$,
\begin{equation}
\label{R_functional}
{\cal R}_0[{\bf f}] = \frac{1}{2}\int\!\!\!\int dtdt' f_i(t){\cal
F}_{ij}(t-t')f_j(t').
\end{equation}
The matrix $\mathcal{F}$ is the inverse of $\varphi_{ij}(t-t')/D$,
\begin{equation}
\label{phi_inverse}
\int dt'{\cal F}_{ij}(t-t')\varphi_{jk}(t'-t'')=D\delta_{ik}\delta(t-t'').
\end{equation}

We are interested in noise realizations that lead to escape, and
therefore ${\bf f}(t)$ largely exceeds its root-mean-square
value. From Eq.~(\ref{P_functional}), the probabilities of such noise
realizations are exponentially small and exponentially strongly depend
on the form of ${\bf f}(t)$. As a consequence, escape trajectories
should form a narrow "tube" centered at an optimal path ${\bf f}_{\rm
opt}(t)$ that maximizes $\mathcal{P}[{\bf f}(t)]$, i.e., minimizes
$\mathcal{R}_0[{\bf f}(t)]$. The minimum of $\mathcal{R}_0$ should be
found with the constraints that (i) the system and noise trajectories,
${\bf q}_{\rm opt}(t)$ and ${\bf f}_{\rm opt}(t)$, are interrelated by
the equation of motion (\ref{eom}), (ii) the path ${\bf q}_{\rm
opt}(t)$ starts in the vicinity of the stable state ${\bf q}_a(t)$ and
ends behind or on the boundary of the basin of attraction to ${\bf
q}_a(t)$, and (iii) the force ${\bf f}_{\rm opt}(t)$ is equal to zero
before the escape event happens and becomes equal to zero once the
system has escaped, so that, as ${\bf f}_{\rm opt}(t)$ decays, it does
not drag the system back to the basin of attraction to ${\bf q}_a$.

As explained in Appendix~A, these conditions lead to boundary
conditions for optimal paths of the form
\begin{eqnarray}
\label{boundary}
&&{\bf q}_{\rm opt}(t)\to \left\{
\begin{array}{l}
{\bf q}_a(t)\;{\rm for}\; t\to -\infty,\\
{\bf q}_b(t)\;{\rm for}\; t\to \infty,
\end{array}\right.\nonumber\\
&&{\bf f}_{\rm opt}(t)\to 0 \;{\rm for}\; t\to \pm\infty
\end{eqnarray}
(note that ${\bf q}_{\rm opt}(t)$ ends on the basin boundary, not on
another attractor.)

The variational problem for optimal paths is reduced to minimizing the
functional
\begin{eqnarray}
\label{var_problem}
{\cal R}[{\bf q},{\bf f}]&={\cal R}_0[{\bf f}] \nonumber\\
&&\!\!\!\!\!\!\!\!\!\!\!\!\!\!\!\!\!\!\!\!\!\!\!\!+\int dt'{\bm\lambda}(t')\cdot[\dot{\bf q}(t')-{\bf K}({\bf q};A,t')-
{\bf f}(t')]
\end{eqnarray}
with boundary conditions (\ref{boundary}). The function ${\bm
\lambda}(t)$ is a Lagrange multiplier. The boundary condition for it
follows from the fact that the system only needs to be driven by a
large force ${\bf f}(t)$ when it moves from the attractor to the basin
boundary. Therefore ${\bm\lambda}(t)\to 0$ for $t\to
\pm\infty$.

It follows from Eqs.~(\ref{boundary}), (\ref{var_problem}) and from
the results of Appendix A that the optimal trajectories ${\bf q}_{\rm
opt}(t), {\bf f}_{\rm opt}(t)$ are instanton-like \cite{Langer}. The
typical duration of motion is given by the relaxation time of the
system $\tr$ and the noise correlation time $t_{\rm corr}$. In stationary systems instantons are
translationally invariant with respect to time, i.e., if ${\bf q}_{\rm
opt}(t), {\bf f}_{\rm opt}(t)$ is a solution, then ${\bf q}_{\rm
opt}(t+\tau), {\bf f}_{\rm opt}(t+\tau)$ is also a solution, for an
arbitrary $\tau$. In contrast, in periodically modulated systems this
is true only for $\tau = \tau_F$.  The instantons are synchronized by
the modulation: generally there is one instanton per period that would
provide a global minimum to ${\cal R}$.

From Eq.~(\ref{P_functional}), we obtain for the escape probability
\begin{equation}
\label{escape_rate}
W\propto\exp(-R/D),\quad R=\min
{\cal R}[{\bf q,f}].
\end{equation}
The activation energy $R$ is equal to the functional ${\cal R}_0[{\bf
f}_{\rm opt}]$ calculated for the optimal noise trajectory for
escape.

For small noise intensity, the escape rate $W\ll \omega_F$. It
periodically depends on time. However, in the small-$D$ limit this
dependence is seen only in the prefactor
\cite{prefactor,Hanggi-00,M&S-01}. Here we are interested in the
exponent, which gives the period-averaged escape rate $\overline
W$. It is equal to the probability of escape over the time $\tau_F$
divided by $\tau_F$

In the general case, the variational problem for the activation energy
can be solved only numerically. Therefore it is particularly important
to find model-independent properties of $R$. So far they have been
found for comparatively weak modulation, where it was
shown that $R$ has a term linear in the modulation amplitude
\cite{LS_chaos}. In this paper we analyze the activation energy $R$ in
a previously unexplored region near a bifurcation point and show that
$R$ displays a nontrivial scaling behavior in this region.


\section{Dynamics near a bifurcation point}


The dynamics near a saddle-node bifurcation point has universal
features related to the occurrence of a slow variable, or a ``soft
mode'' \cite{Guckenheimer}.  For periodically modulated systems,
closeness to the bifurcation point in the parameter space usually
implies that the merging states are close to each other in phase space
throughout the modulation period.

If the modulation frequency $\omega_F$ is small compared to the
reciprocal relaxation time in the absence of modulation $1/\tr^{(0)}$,
there emerges a situation where the stable and unstable states come
close to each other only for a portion of a period. During this time,
the system behaves as if it were close to a true bifurcation point.
Then it is possible to single out a slow variable that controls
the system dynamics. Escape from a metastable
state is most probable when ${\bf q}_a(t)$ and ${\bf q}_b(t)$
are closest to each other.

On the other hand, if the modulation frequency $\omega_F\agt
1/\tr^{(0)}$, near the bifurcation point the states ${\bf q}_a(t)$ and
${\bf q}_b(t)$ are close to each other throughout the modulation
period. Then escape events should be no longer synchronized by the
modulation.

Because the dynamics near a bifurcation point is slow, the system filters
out high-frequency components of the noise. As a result, the noise
becomes effectively $\delta$-correlated (we will not consider here the
situation where the noise power spectrum has singular features at high
frequencies). The reduction to one slow variable driven by white noise
can be done directly in the equations of motion. For slow modulation
($\omega_F\tr^{(0)}\ll 1$), this reduction is local in time (see Appendix~B), otherwise
it has to be done globally over the cycle (see Appendix~C). Alternatively, the
dimensionality reduction can be done directly in the variational
problem for the optimal escape path (see Appendix~A).

\subsection{The adiabatic approximation}

In the limit of slow modulation, where the period of the field
$\tau_F$ is large compared to the system relaxation time $\tr^{(0)}$, a
convenient starting point of the analysis is the adiabatic
approximation.  The adiabatic periodic states of the system ${\bf
q}_{\rm per}^{\rm ad}$ are given by the equation
\begin{equation}
\label{ad_periodic_}
{\bf K}({\bf q}_{\rm per}^{\rm ad};A,t)=0,
\end{equation}
which is obtained by disregarding $\dot{\bf q}$ and the noise ${\bf
f}$ in the equation of motion (\ref{eom}).

The adiabatic stable state (attractor)
${\bf q}_a^{\rm ad} \equiv {\bf q}_a^{\rm ad}(t)$ is the solution
${\bf q}_{\rm per}^{\rm ad}$ for which the real parts of the
eigenvalues of the matrix $\hat{\mu}=\left(\partial K_i/\partial
q_j\right)$ are all negative. These eigenvalues give the
``instantaneous'' relaxation rates, for a given phase of the
modulation $\phi=\omega_Ft$. For the periodic adiabatic saddle-type
state ${\bf q}_b^{\rm ad}(t)$ one of the eigenvalues of $\hat{\mu}$
has a positive real part.

In the adiabatic approximation, the saddle-node bifurcation occurs in
the following way. At the critical value of the control parameter $A =
A_c^{\rm ad}$, the periodic trajectories ${\bf q}_a^{\rm ad}(t)$ and
${\bf q}_b^{\rm ad}(t)$ given by (\ref{ad_periodic_}) merge, but it
happens only once per period. One can picture it by looking at
Fig.~\ref{fig:barrier}~(b) and imagining that the states ${\bf q}_a(t)$ and
${\bf q}_b(t)$ touch each other. We set the corresponding instant of time
equal to $t=0$ (or $t=n\tau_F$), i.e., we assume that ${\bf q}_a^{\rm
ad}(0)={\bf q}_b^{\rm ad}(0)$ for $A = A_c^{\rm ad}$. Additionally, we
set ${\bf q}_a^{\rm ad}(0)={\bf q}_b^{\rm ad}(0)={\bf 0}$.

At the adiabatic bifurcation point $A=A_c^{\rm ad}, t=0$ one of the
eigenvalues $\mu_1$ of the matrix $\hat{\mu}$ is equal to
zero, whereas all other eigenvalues $\mu_{i>1}$ have {\em
negative} real parts. The adiabatic approximation means that
Re~$\mu_{i>1} \gg \omega_F$, or equivalently, that the relaxation
time $ \max |{\rm Re}\,\mu_{i>1}|^{-1}$ is small compared to
$\tau_F$. This relaxation time is typically of the order of
$\tr^{(0)}$.

We now write the dynamical variables ${\bf q}$ in the basis of the
right eigenvectors of the matrix $\hat{\mu}$ at the bifurcation point
and expand ${\bf K}$ in the equation of motion (\ref{eom}) near the
bifurcation point ${\bf q}={\bf 0}, t=0, A=A_c^{\rm ad}$. As shown in
Appendix~B, the motion described by the variable $q_1$ is much slower
than the motion described by the variables $q_{i>1}$. Over the time
$\sim \tr^{(0)}$ they ``adjust'' to the value of $q_1$, i.e., they
follow $q_1$ adiabatically. The variable $q_1$ is the soft mode. It
satisfies the equation of motion
\begin{eqnarray}
\label{q_1_ad_}
&&\dot q_1 = K(q_1;A,t) + f_1(t),\\
&&K=\alpha q_1^2 + \beta \,\delta\!A^{\rm ad} - \alpha\gamma^2 (\omega_Ft)^2.\nonumber
\end{eqnarray}
Here $\delta\!A^{\rm ad}=A-A_c^{\rm ad}$; the parameters
$\alpha,\beta,\gamma$ are expressed in terms of the derivatives of
${\bf K}$ at the adiabatic bifurcation point and are given by
Eqs.~(\ref{q_1_ad}), (\ref{gamma^2}).

The stable and unstable adiabatic periodic states in the absence of
noise exist for $\alpha\beta\,\delta\!A^{\rm ad}<0$.  For concreteness
and without loss of generality we set $\alpha>0$.  For small
$|\omega_Ft|$ the adiabatic states can be found by setting
$K((q_1)_{a,b}^{\rm ad};A,t)=0$. This gives
%
\[(q_1)_{a,b}^{\rm ad} = \mp
(2\alpha\tr^{\rm ad})^{-1},\]
%
where $\tr^{\rm ad}$ is the instantaneous adiabatic relaxation
time. It is given by $|\partial K/\partial q_1|^{-1}$ evaluated for
 $q_1=(q_1)_a^{\rm ad}$,
\begin{equation}
\label{tr_ad}
\tr^{\rm ad}=\frac{1}{2}\left[-\alpha\beta\delta\!A^{\rm ad}+(\alpha\gamma\omega_Ft)^2\right]^{-1/2}
\end{equation}
[the explicit expression for $(q_1)_{a,b}^{\rm ad}(t)$ is given by
Eq.~(\ref{ad_states})].

The applicability of the adiabatic approximation requires not only
that $\omega_F\tr^{\rm ad}\ll 1$, but also $|\partial\tr^{\rm
ad}/\partial t|\ll 1$, otherwise the system cannot follow the
modulation.  From Eq.~(\ref{tr_ad}), near the bifurcation point the
time dependence of $\tr^{\rm ad}$ is pronounced, so that
%
\[\max|\partial\tr^{\rm ad}/\partial t| =
3^{-3/2}\gamma\omega_F/|\beta\delta\!A^{\rm ad}|\gg\omega_F\tr^{\rm ad}.
\]
%
Therefore the inequality $|\partial\tr^{\rm ad}/\partial t| \ll 1$ is
much stronger than $\omega_F\tr^{\rm ad}\ll 1$. It holds if
\begin{equation}
\label{ad_condition_}
\tr^{\rm ad} \ll t_l, \qquad t_l = (\alpha\gamma\omega_F)^{-1/2},
\end{equation}
i.e., $\omega_F\ll |\beta\delta\!A^{\rm ad}|/\gamma$. The time $t_l$
sets a new dynamical time scale that restricts the range of validity
of the adiabatic approximation. It imposes an upper bound on the
relaxation time of a periodically driven system where this approximation still applies.


\subsection{Locally nonadiabatic regime}

As $A$ approaches $A_c^{\rm ad}$, the criterion (\ref{ad_condition_})
is violated. The periodic stable and unstable states are pressed
against each other. Since they cannot cross, they become distorted, as
shown in Fig.~\ref{fig:barrier}c. Ultimately they merge, but along a
line rather than at a point. From Eq.~(\ref{q_1_ad_}) one can see that
this is just a straight line, which is described by the equation
\begin{equation}
\label{q_critical}
q_{1c}(t)=\gamma\omega_Ft.
\end{equation}

Eqs.~(\ref{q_1_ad_}), (\ref{q_critical}) define the new nonadiabatic
bifurcational value of the modulation amplitude for slow driving,
\begin{equation}
\label{weak_bif}
A_c^{\rm sl} =  A_c^{\rm ad} + \beta^{-1}\gamma\omega_F.
\end{equation}
The corrections to $A_c^{\rm sl}$ of higher order in $\omega_F$ are
discussed in Sec.~III~C and obtained explicitly in Sec.~V.

It is convenient to change in Eq.~(\ref{q_1_ad_}) to dimensionless
variables $Q=\alpha t_lq_1$ and $\tau=t/t_l$, and to introduce the
control parameter
\begin{equation}
\label{eta_def}
 \eta =\beta(\gamma\omega_F)^{-1}(A_c^{\rm sl}-A).
\end{equation}
This is the only parameter of a slowly driven system. It describes
both adiabatic and nonadiabatic behavior and gives the reduced distance
to the bifurcation point.

The equation of motion for the reduced variable
takes the form
\begin{equation}
\label{nonad_eom}
{dQ\over d\tau} = G(Q,\eta, \tau) + \tilde f(\tau),\; G=Q^2 -\tau^2 +1-\eta.
\end{equation}

The function $\tilde f(\tau)= (\gamma\om)^{-1}f_1(t)$ in
Eq.~(\ref{nonad_eom}) describes reduced noise. It is effectively
$\delta$-correlated on the slow-time scale,
$\langle\tilde{f}(\tau)\tilde{f}(\tau')\rangle=2\tilde{D}\delta(\tau-\tau')$,
as explained in Appendix~B.  The noise intensity $\tilde{D}$ is given
by Eq.~(\ref{tilde_D}).

The stable and unstable states $Q_{a,b}(\tau)$ are given by the equation
 $dQ/d\tau=G$.
In the absence of noise this equation has symmetry $Q\to -Q,
\tau \to -\tau$. As a consequence, the stable and unstable states are
antisymmetric, $Q_b(\tau)= -Q_a(-\tau)$. Therefore it suffices to find
only $Q_a(\tau)$.

We start with the adiabatic approximation. It applies for $\eta\gg 1$.
The adiabatic stable and unstable states in the reduced variables are
given by the equation $G=0$ and have the form
\begin{equation}
\label{ad_states_}
Q^{\rm ad}_{a,b} = \mp\left[\tau^2+(\eta-1)^2\right]^{1/2}.
\end{equation}
Each of these states is symmetric with respect to $\tau=0$, where
they are closest to each other.  The adiabatic bifurcation point is
$\eta=1$, which corresponds to $A=A_c^{\rm ad}$.

The region $\eta\lesssim 1$ is nonadiabatic, and $\eta=0$ (or
$A=A_c^{\rm sl}$) is the nonadiabatic bifurcation point for slow
driving.  At this point $Q_{a,b}(\tau)$ merge into the straight line
$Q_c(\tau)=\tau$.

Close to the nonadiabatic bifurcation point, where $\eta \ll 1$, one
can find $Q_{a,b}(\tau)$ by perturbation theory in the whole range
$-\infty < \tau < \vert\ln\eta\vert^{1/2}/2$. The linearized equation
for the difference $\delta Q(\tau) = Q_a(\tau)-\tau$ has a simple form
%
${d\delta Q\over d\tau}=2\tau \,\delta Q +\eta$.
%
By solving it we obtain
\begin{eqnarray}
\label{weak_q_ab}
&Q_a(\tau)=- Q_b(-\tau)\approx \tau -\eta\int\nolimits_{-\infty}^{\tau}
d\tau_1\,e^{\tau^2-\tau_1^2}.
\end{eqnarray}

In the region of large negative $\tau$ the function $Q_a(\tau)=-
Q_b(-\tau)$ has a simple form $Q_a(\tau)\approx
-\tau-\eta(2\tau)^{-1}$. The states $Q_a$ and $Q_b$ are closest to
each other, with separation $\sim \eta$, in the range $|\tau|<
\vert\ln \eta\vert^{1/2}/2$. The interstate separation decreases as
$\eta$ approaches the bifurcational value $\eta =0$. At the same time,
the range of $\tau$ where $Q_a$ and $Q_b$ are close to each other
increases with decreasing $\eta$.

\begin{figure}[ht]
\includegraphics[width=2.5in]{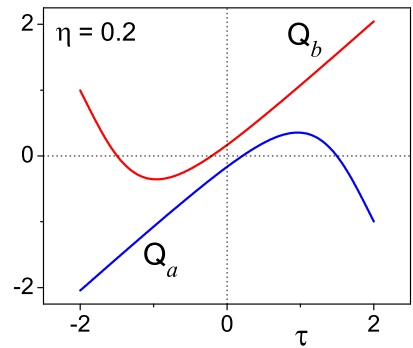}
\caption{Nonadiabatic stable and unstable states $Q_{a}(\tau)$ and
$Q_{b}(\tau)=-Q_a(-\tau)$ for slow modulation as given by the equation
$dQ/d\tau = G(Q,\eta,\tau)$ for $\eta=0.2$. The functions
$Q_{a,b}(\tau)$ are strongly asymmetric, in contrast to the adiabatic
states (\protect\ref{ad_states_}) which are even functions of $\tau$.}
\label{fig:nonad}
\end{figure}

As $\tau$ increases beyond $\approx \vert\ln\eta\vert^{1/2}/2$, there
occurs a sharp crossover from the nearly linear in $\tau$ solution for
$Q_a(\tau)$ (\ref{weak_q_ab}) to the adiabatic solution
(\ref{ad_states_}) $Q_a\propto -\tau$.  The functions $Q_{a,b}(\tau)$
for a specific value of $\eta$ are shown in Fig.~\ref{fig:nonad}.

The interval of the real time $|t|\lesssim \tr =t_l|\ln\eta|^{1/2}$, where the
states $Q_{a,b}$ are nearly linear in $t$, should be much smaller than
$1/\omega_F$ in order for Eq.~(\ref{q_1_ad_}) to apply.  This imposes
a restriction on $\eta$,
\begin{equation}
\label{weak_limit}
\eta \gg\exp(-C|\alpha|\gamma/\omega_F),\quad C\sim 1.
\end{equation}
For smaller $\eta\propto |A_c^{\rm sl}-A|$ the local approximation,
where the coefficients are expanded about the adiabatic bifurcation
point, no longer applies.  The relaxation time $t_r$ becomes
comparable to the modulation period. It follows from
Eq.~(\ref{weak_limit}), however, that for low frequencies the local
approximation is extremely good.

On the whole, the locally nonadiabatic regime is limited in $\eta$ by
the condition $\eta \alt 1$ and by Eq.~(\ref{weak_limit}). The width of the
amplitude range $A_c^{\rm sl} - A$ imposed by the first condition linearly
increases with the field frequency, in the approximation
(\ref{weak_bif}). Therefore locally-nonadiabatic critical behavior is
more pronounced for higher frequencies. However, the appropriate
frequency range is limited from above by the condition
(\ref{weak_limit}). For higher $\omega_F$ there should occur a
crossover to a fully nonlocal picture, which is discussed in the next
section.


\subsection{Fast-oscillating field}

Sufficiently close to the ''true'' critical value of the modulation
amplitude $A_c$, the relaxation time of the system becomes large
compared to the modulation period even if $\omega_F\tr^{(0)}\lesssim
1$ far from the bifurcation point. The inequality $\omega_F\tr \gg 1$
defines the third region, in addition to the adiabatic and locally
nonadiabatic, where we have analyzed the dynamics near a bifurcation
point. The analysis of this region is simplified by the fact that here
the modulating field is effectively fast-oscillating.

\begin{figure}[ht]
\includegraphics[width=2.5in]{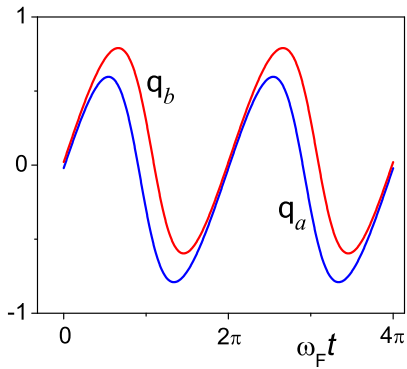}
\caption{The stable and unstable states, ${\bf q}_a(t)$ and ${\bf
q}_b(t)$, close to the bifurcation point. For $\omega_F\tr \gg 1$ the
states are close to each other throughout the modulation period. The
figure refers to a one-dimensional overdamped particle in a potential
$U(q,t)={1\over 4}q -{1\over 3}q^3 -Aq\cos(\omega_Ft)$ for
$(A_c-A)/A_c \approx 0.01$. The modulation is comparatively slow,
$\omega_F\tr^{\rm (0)}=1$, but for chosen $A$ the relaxation time
becomes long, $\omega_F\tr \approx 9.8$.}
\label{fig:fast}
\end{figure}

For $\omega_F\tr \gg 1$, near the bifurcation point the periodic
stable and unstable states ${\bf q}_a(t)$ and ${\bf q}_b(t)$ stay
close to each other throughout the cycle, see Fig.~\ref{fig:fast}.
For $A=A_c$, they coalesce into a periodic
critical cycle ${\bf q}_c(t)={\bf q}_c(t+\tau_F)$.  When $A$ is close to
$A_c$ and ${\bf q}$ is close to ${\bf q}_c$, we can simplify the
equations of motion (\ref{eom})
by expanding the function ${\bf K}$ in $\delta {\bf q} = {\bf q}- {\bf
q}_c$ and $\delta\!A=A-A_c$ (cf. Ref.~\onlinecite{Guckenheimer}),
\begin{equation}
\label{near_bifurcation}
\delta \dot{\bf q} = \hat\mu\delta {\bf q} + \frac{1}{2}
(\delta {\bf q}\cdot\bm{\nabla})^2{\bf K} +\delta\!A{\partial\over
\partial A}{\bf K} + {\bf f}(t).
\end{equation}
Here $\mu_{ij}\equiv \mu_{ij}(t)= \partial K_i/\partial q_j$. All
derivatives of ${\bf K}$ are now evaluated for $A=A_c$ and ${\bf q}=
{\bf q}_c(t)$. Therefore all coefficients in
Eq.~(\ref{near_bifurcation}) are periodic functions of time.

If initially the system is close to ${\bf q}_c(t)$, its distance from
${\bf q}_c(t)$ will oscillate with frequency $\omega_F$ and with an
amplitude that slowly varies over the period $\tau_F$. This amplitude
is a slow variable, $Q^{\rm sm}(t)$. The equation for $Q^{\rm sm}(t)$ can be obtained
by an appropriate averaging method explained in Appendix~C. After
rescaling to dimensionless coordinate $Q\propto Q^{\rm sm}$ and time
$\tau\propto t$, see Eq.~(\ref{fast_Q}), this equation takes a form
which is similar to Eq.~(\ref{nonad_eom}),
\begin{eqnarray}
\label{slow_reduced}
&&{dQ\over d\tau}= G(Q,\eta) + \tilde f(\tau),\\
&& G= Q^2-\eta, \qquad
\eta = \beta'(A_c-A) \nonumber
\end{eqnarray}
[in contrast to Eq.~(\ref{nonad_eom}), the function $G$ here is
independent of time]. The coefficient $\beta'$ is given by
Eq.~(\ref{coefficients}).

The parameter $\eta$ in Eq.~(\ref{slow_reduced}) is the scaled
distance to the bifurcation point. The stationary states
$Q_{a,b}=\mp\eta^{1/2}$ exist for $\eta > 0$. They merge for $\eta=0$.
The noise $\tilde f(\tau)$ is effectively
white on the time scale that largely exceeds $\omega_F^{-1}$ and the
noise correlation time $t_{\rm corr}$. Its intensity $\tilde D$ is
given by Eq.~(\ref{noise_fast}).

The results of this Section and Appendix C refer to the case
$\omega_F\tr \gg 1$, but arbitrary $\omega_F\tr^{(0}$.
Therefore the problem is different from the standard problem
of slow motion in a fast-oscillating field \cite{LL-Mechanics}, where
of interest is the smooth term in the oscillating coordinate. In
contrast, here we are interested in the slowly varying oscillation
amplitude.  If $\omega_F\tr^{(0)} \gg 1$, a
transition to slow and fast variables can be made already in the
original equation of motion (\ref{eom}), by separating ${\bf q}$ into
slow and fast oscillating parts. The equation for the slow part near
the bifurcation point will again have the form (\ref{slow_reduced}),
but the expressions (\ref{coefficients}) for $\alpha',\beta'$ will be
simplified [in particular, the factor $\kappa_{11}$ in
Eq.~(\ref{coefficients}) will be equal to one].

\subsection{Connection to the locally nonadiabatic regime}

Eq.~(\ref{near_bifurcation}) allows us to look from a different
perspective at the locally nonadiabatic regime that emerges for $\omega_F\tr\ll 1$.
 In contrast to the approach of Sec.~III~B,
where the starting point was the adiabatic approximation, here we will
assume that $A$ is close to the true bifurcational value of the
amplitude $A_c$ and that ${\bf q}(t)$ is close to the critical cycle
${\bf q}_c(t)$, at least for a part of the period $\tau_F$.

For $\omega_F\tr\ll 1$, one can think of a local in time description
of the dynamics near the cycle ${\bf q}_c(t)$. From
Eq.~(\ref{near_bifurcation}), this dynamics is determined by the
eigenvalues $\mu_{\nu}(t)$ of the matrix $\hat\mu(t)$. In contrast to
the analysis of Sec.~III~A, we consider here the matrix $\hat\mu$
calculated for the critical cycle ${\bf q}_c(t)$ rather than the two
similar matrices calculated separately for the adiabatic stable and
unstable states.

For much of the driving period the real parts of $\mu_{\nu}(t)$ are
all large, $|$Re~$\mu_{\nu}|\sim 1/\tr^{(0)}\gg \omega_F$.  Then, when
the system is in the stable state, it follows the field
adiabatically. The adiabaticity is broken where one of the
eigenvalues, say $\mu_1(t)$, goes through zero. As we will see, at
this time the stable and unstable states are most close to each
other. We set the time when it happens equal to zero, i.e.,
$\mu_1(0)=0$.

For small $|t|\ll \tau_F$ the analysis of the system dynamics is in
many respects similar to that in Sec.~III~B and Appendix~C. First,
$\delta {\bf q}(t)$ in Eq.~(\ref{near_bifurcation}) is written as
$\sum\nolimits_{\nu}\delta q_{\nu}{\bf e}_{\nu}(0)$, where ${\bf
e}_{\nu}(0)$ are the right eigenvectors of the matrix
$\hat\mu(0)$. The component $\delta q_1$ of $\delta {\bf q}$ along the
eigenvector ${\bf e}_1(0)$ of $\hat\mu(0)$ will be the slow variable,
or the soft mode.

The matrix
$\hat\mu(t)$ can be expanded about $t=0$ for small $|t|$,
\begin{equation}
\label{lambda_expansion}
\hat\mu(t)\approx \hat\mu(0)
+ \dot{\hat{\mu}}(0)t,
\end{equation}
where the time derivative is taken for $t=0$. This derivative is
small, its matrix elements on the eigenvectors $\bar{\bf
e}_{\nu}(0), {\bf e}_{\nu'}(0), $ are
$|\dot{\mu}_{\nu\nu'}| \sim \omega_F/\tr^{(0)}\ll
\bigl(\tr^{(0)}\bigr)^{-2}$ [here, $\bar{\bf e}_{\nu}(0)$ are
the left eigenvectors of the matrix $\hat\mu(0)$].

With (\ref{lambda_expansion}), Eq.~(\ref{near_bifurcation}) can
be solved for the "fast" components $\delta q_{\nu>1}$. Over a short
time $\sim \tr^{(0)}$ they approach their quasistationary values for given
$\delta q_1$. Those are small, of order $\delta q_1^2, \delta\!A,
\delta q_1 \omega_Ft$, and follow $\delta q_1$ adiabatically.
Noise-induced fluctuations of $\delta q_{\nu>1}$ about the
quasistationary values are also small for small noise intensity.
Therefore the effect of $\delta q_{\nu>1}$ on the dynamics of $\delta
q_1$ can be disregarded.

The equation of motion for $\delta q_1$ is of the form
\begin{eqnarray}
\label{local_lambda}
\delta \dot q_1 &=& \dot\mu_{1} t\delta q_1 +\alpha\delta q_1^2 +
\beta \delta\!A + f_1(t),\\
f_1(t)&=& \bar{\bf e}_1(0)\cdot{\bf f}(t),\qquad  \dot\mu_{1} =
\bar{\bf e}_1(0)\cdot\dot{\hat\mu}(0){\bf e}_1(0).\nonumber
\end{eqnarray}
Here, $\alpha=(1/2) \bigl({\bf
e}_1(0)\cdot{\bm\nabla}\bigr)^2K_1,\, \beta = \partial
K_1/\partial A$, with $K_1=\bar{\bf e}_1(0)\cdot{\bf K}$ being
now the component of ${\bf K}$ in the direction ${\bf e}_1(0)$.
All derivatives of ${\bf K}$ are calculated on the critical cycle
${\bf q}_c(t)$ for
$t=0$. Because $|\dot\mu_{1}| \sim \omega_F/\tr^{(0)}$ is small,
relaxation of $\delta q_1$ is slow compared to relaxation of $\delta
q_{\nu>1}$, for typical $|t|\ll \tau_F$.

Eq.~(\ref{local_lambda}) describes the stable and unstable states of
the original equation of motion (\ref{eom}) in the region $|t|\ll
\tau_F$. It is seen that these states, $(\delta q_1)_{a}$ and $(\delta
q_1)_{b}$, exist provided
\begin{equation}
\label{weak_lambda_cond}
\dot\mu_{1} > 0,\quad \alpha\beta\,\delta \!A < 0.
\end{equation}
In this range Eq.~(\ref{local_lambda}) is equivalent to
Eq.~(\ref{nonad_eom}). This can be seen if, on the one hand,
Eq.~(\ref{nonad_eom}) is written for the deviation $\delta Q =
Q-\tau$ of $Q$ from its value on the critical cycle $Q_c =\tau$,
and on the other hand, in Eq.~(\ref{local_lambda}) one changes to
scaled variables $\delta Q = \alpha(2/\dot\mu_1)^{1/2}\delta
q_1$ and $\tau
=(\dot\mu_{1}/2)^{1/2}t$. The control parameter $\eta$ in
(\ref{nonad_eom}) becomes
\begin{equation}
\label{xi_renormalized}
\eta = -2\alpha\beta\dot\mu_{1}^{-1}\delta\!A.
\end{equation}
The analysis of Eq.~(\ref{nonad_eom}) then applies to
Eq.~(\ref{local_lambda}). In particular, the statement in the
beginning of this subsection that the stable and unstable states are
at their closest for $t=0$ is an immediate consequence of the explicit
expression for these states (\ref{weak_q_ab}).

There is an important difference between this approach and the
approach of Sec.~III~B. Because here we do not start from the
adiabatic approximation, we formally have not specified how small is
the difference between the critical amplitude $A_c$ and its adiabatic
value $A_c^{\rm ad}$. In Eq.~(\ref{weak_bif}) we only obtained the
linear in $\omega_F\tr^{(0)}$ term in $A_c-A_c^{\rm ad}$. In general,
$A_c-A_c^{\rm ad}$ has also higher-order terms. They can be obtained
by taking into account the dependence of the coefficients $\alpha,
\beta, \gamma$ in Eq.~(\ref{q_1_ad}) on $A$, which was previously
disregarded. This is illustrated for a particular model in Sec.~V. It
is for the renormalized critical amplitude, i.e., for the control
parameter given by Eq.~(\ref{xi_renormalized}), that there holds the
exponential limit (\ref{weak_limit}) on the range where the local
nonadiabatic approximation applies. The inequality (\ref{weak_limit})
indicates that, for small frequency, the critical amplitude found from
the local theory is exponentially close to the exact $A_c$. This is
confirmed by numerical calculations for a model discussed in Sec.~V.

\section{Activation energy of escape}

It follows from the results of Sec.~III that, near a bifurcation
point, a periodically driven system has a soft mode $Q$, and the noise
that drives this mode is effectively white. The equation of motion is
of the form $dQ/d\tau = G + f(\tau)$ (\ref{nonad_eom}), where the
function $G$ is given by $G=Q^2 + 1-\eta - \tau^2$ for $\omega_F\tr \ll
1$ [cf. Eq.~(\ref{nonad_eom})] and $G=Q^2-\eta$ for $\omega_F\tr \gg
1$ [cf. Eq.~(\ref{slow_reduced})]. The intensity $\tilde D$ of the noise
$f(\tau)$ has the form (\ref{tilde_D}) and (\ref{noise_fast}) in these
two cases.

For a white-noise driven system, the variational problem
(\ref{var_problem}), (\ref{escape_rate}) of calculating the
period-averaged rate of activated escape $\overline W$ can be written in the
form
\begin{eqnarray}
\label{R_1d}
&&\overline W={\rm const}\times \exp(-\tilde R/\tilde D), \quad \tilde
R=\min\tilde {\cal R}[Q],\\
&&\tilde {\cal R} = \int d\tau\,
L\left(Q,{dQ\over d\tau},\tau\right), \quad L={1\over 4}\left({dQ\over
d\tau}-G\right)^2.\nonumber
\end{eqnarray}
(cf. Appendix~A).
In contrast to the standard formulation
\cite{Freidlin84}, the function $G$ here may depend on time $\tau$ and
is not time-periodic, in the actual range of $\tau$. The minimization
is carried out over the paths $Q(\tau)$ that start at $Q_a(\tau)$ for
$\tau\to-\infty$ and end at $Q_b(\tau)$ for $\tau\to+\infty$, with time-dependent
stable and unstable states $Q_{a,b}(\tau)$.
The non-stationarity emerges for slow modulation, where $\omega_F\tr \ll
1$, and is related to the assumptions that (i) escape is most likely to
occur during a portion of the period where the states $Q_{a,b}$ are
close, and (ii) the duration of motion along the optimal escape path
$Q_{\rm opt}(\tau)$ is much less than the modulation period.

We have solved the variational problem using the Hamilton equations of
motion for $Q$ and $P=\partial L/\partial(dQ/d\tau)$,
\begin{equation}
\label{eqn_H}
{\partial Q\over \partial \tau} =2P+G,\qquad
{\partial P\over \partial\tau}=-P\,{\partial{G}\over\partial{Q}}.
\end{equation}
We then verified the assumptions made in obtaining Eqs.~(\ref{R_1d}),
(\ref{eqn_H}).

Equations (\ref{eqn_H}) were solved numerically. This
was done by choosing the initial conditions on the optimal path close
to $Q_a(\tau)$ with large but finite negative $\tau$. In this range
Eqs.~(\ref{eqn_H}) can be linearized in $Q-Q_a(\tau)$. On the solution
that goes away from $Q_a$ the momentum $P$ is linear in $Q-Q_a$,
\begin{eqnarray}
\label{initial}
&&P\approx
[Q-Q_a(\tau)]/\sigma_a^2(\tau)\\
&&\sigma_a^2(\tau)=2\int\nolimits_{-\infty}^{\tau}d\tau'\,
\exp\left[4\int\nolimits_{\tau'}^{\tau}d\tau''\,Q_a(\tau'')\right]\nonumber
\end{eqnarray}
We used the shooting method:
we sought such initial $Q-Q_a(\tau)$ for given $\tau$ that the
trajectory approaches $Q_b(\tau)$ for large $\tau$,
cf.~Ref.\onlinecite{LS_chaos}.

Numerical results for the activation energy in the whole range of slow
driving, where $\omega_F\tr \ll 1$, are shown below in
Figs.~\ref{fig:adcorr}, \ref{fig:true_bif_states} on linear and
logarithmic scales, respectively.  Note that the activation energy is
a function of a single control parameter $\eta\propto A_c-A$, and in
this sense the results are universal, i.e. system-independent. In the
rest of this Section we discuss analytical results and compare them
with the numerical results.

\subsection{Activation energy in the adiabatic approximation}

The adiabatic regime applies when the driving is slow, $\omega_F\tr
\ll 1$, and the system is sufficiently far from the bifurcation point,
so that $\tr\ll t_l$ [cf. Eq.~(\ref{ad_condition_})] or equivalently
$|\delta\!A^{\rm ad}|\equiv |A_c^{\rm ad} -A|\gg
\omega_F|\gamma/\beta|$. Respectively, the dimensionless control
parameter $\eta \propto A_c-A$ is large, $\eta-1 \gg 1$,
[cf. Eq.~(\ref{eta_def}); we note that the actual parameter in the
adiabatic range is not $\eta$ but $\eta - 1$]. In this case we expect
that escape occurs when the adiabatic states (\ref{ad_states_})
%
 are closest to each
other, i.e., for $\tau = 0$. Then, in the first approximation, the
term $\tau^2$ in the function $G$ in Eq.~(\ref{nonad_eom}) can be
disregarded, and $G$ becomes
\begin{equation}
\label{M_ad}
G^{\rm ad} = Q^2+1-\eta.
\end{equation}

The solution of Eq.~(\ref{eqn_H}) with $G$ of the
form (\ref{M_ad}) and with boundary conditions $Q(\tau)\to \mp(\eta -
1)^{1/2}$ for $\tau\to \mp\infty$ is well known. It is an instanton (kink)
\begin{equation}
\label{q_opt_ad}
Q_{\rm opt}^{\rm ad}(\tau,\tau_0)
=(\eta-1)^{1/2}\tanh[(\eta-1)^{1/2}(\tau-\tau_0)] \nonumber
\end{equation}
centered at an arbitrary $\tau_0$.

The characteristic duration of motion along the path $Q_{\rm opt}^{\rm
ad}(\tau)$ in dimensionless time is $\Delta\tau \sim (\eta
-1)^{-1/2}$, which corresponds to $\Delta t \sim \tr$ in real
time. Since $\Delta \tau \ll (\eta -1)^{1/2}$, the term $\tau^2$ in
the function $G$ [Eq.~(\ref{nonad_eom})] can be disregarded compared
to $\eta-1$, which justifies replacing $G$ with $G^{\rm ad}$ as long
as $|\tau_0|\ll (\eta-1)^{1/2}$.

The activation energy (\ref{R_1d}) calculated along the path $Q_{\rm
opt}^{\rm ad}$ is
\begin{equation}
\label{R_ad}
\tilde R^{\rm ad} = {4\over 3}(\eta-1)^{3/2}
\propto (A^{\rm ad}_{c}-A)^{3/2}.
\end{equation}
This equation shows that the activation energy of escape scales with the
distance to the bifurcation point as $(A_c-A)^{\xi}$ with
$\xi=3/2$ in the adiabatic region.

\subsection{Nonadiabatic correction to the activation energy}

We now consider the lowest-order correction to the adiabatic
activation energy. Two factors have to be taken into account. First is
that, because of the nonzero duration of motion along the escape path
$\Delta\tau$, the equilibrium states $Q_{a,b}^{\rm ad}(\tau)$ change,
which was disregarded in the analysis of Sec.~IV~A. However, the
corresponding correction to $\tilde R$ is exponentially small. Indeed,
if we consider $\tilde{\cal R}$ as a function of the end point $Q$ on
the optimal path, we have $|\partial \tilde{\cal R}/\partial Q| =
|P|$, where $P$ is the momentum on the optimal path. For an
instantonic solution, the momentum goes to zero as $Q_{\rm opt}\to
Q_{a,b}$, see Eq.~(\ref{initial}). Therefore a small change of
$Q_{a,b}$ affects the activation energy very weakly.

The major nonadiabatic correction to $\tilde R$ comes from the
time-dependent term in $G = G^{\rm ad}-\tau^2$ [cf.
Eq.~(\ref{nonad_eom})]. This term lifts the time invariance of the
instanton $Q_{\rm opt}(\tau,\tau_0)$ with respect to $\tau_0$,

To first order in $\tau^2$, i.e., to lowest order in $(\eta-1)^{-2}$, the
correction $\delta \tilde R$ can be found from Eq.~(\ref{R_1d}) by
integrating the term $\tau^2$ along the zeroth order trajectory
$Q_{\rm opt}^{\rm ad}(\tau,\tau_0)$,
\begin{equation}
\label{correction}
\delta \tilde R=\min_{\tau_0}\int d\tau'\,{d{Q}_{\rm opt}^{\rm
ad}(\tau',\tau_0)\over d\tau'}\,\tau'^2.
\end{equation}
(we used that $dQ_{\rm opt}^{\rm ad}/d\tau = -G^{\rm ad}$).
 Minimization here is done over $\tau_0$, the center of the
instanton. It is necessary because $\tilde R$ is the absolute
minimum of the functional $\tilde{\cal R}$.

A direct calculation shows that the minimum of $\delta \tilde R$ is reached
for $\tau_0=0$, and
\begin{equation}
\label{R_adcorr}
\delta\tilde R = {\pi^2\over 6}(\eta-1)^{-1/2}.
\end{equation}
The correction $\delta\tilde{R}$ rapidly falls off with increasing $\eta
-1$. On the other hand, as $\eta$ decreases and becomes $\sim 1$, the
term $\delta\tilde R$ increases very fast, which
indicates a breakdown of the adiabatic theory in this region.

\begin{figure}[ht]
\includegraphics[width=2.8in]{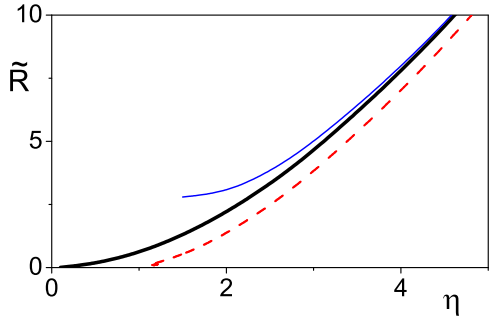}
\caption{The activation energy $\tilde R$ vs. $\eta\propto A_c-A$ for
slow driving, $\omega_F\tr \ll 1$.  The thick solid line shows the
numerical solution of Eq.~(\protect\ref{R_1d}).  The dashed line is the
adiabatic activation energy (\protect\ref{R_ad}), $\tilde R^{\rm
ad}\propto (\eta-1)^{3/2}$. The thin solid line shows the corrected adiabatic
activation energy $\tilde R^{\rm ad}+\delta\tilde R$. It is close
to the numerical result for $\eta \agt 3$. The
correction $\delta\tilde R$ diverges at the adiabatic bifurcation
point $\eta=1$.}
\label{fig:adcorr}
\end{figure}

The analytical results in the adiabatic region are compared with the
numerical solution for the activation energy $\tilde R$ in
Fig.~\ref{fig:adcorr}. The corrected adiabatic theory works well in
the whole range where the control parameter $\eta\agt 3$, but for
smaller $\eta$ nonadiabatic effects are significant and have to be
taken into account in a nonperturbative way.

\subsection{Activation energy in the locally nonadiabatic region}

Standard techniques do not allow to solve equations (\ref{eqn_H})
for the optimal path analytically in the general case $\eta \sim
1$.  This is because the function $G$ in Eq.~(\ref{eqn_H})
explicitly depends on time $\tau$. However, a solution can be
obtained close to the bifurcation point, where $\eta \propto
A_c-A$ is small, but not exponentially small,
cf.~Eq.~(\ref{weak_limit}).

Unusually for an instanton-type problem, and because of the strong
time dependence of $G$, the optimal path can be found by {\it
linearizing} the equations of motion (\ref{eqn_H}) about the critical
state $Q_c=\tau$.  This gives
\begin{equation}
\delta\dot Q=2\tau \delta Q -\eta + 2P, \quad \dot{P}=-2P\tau,
\end{equation}
where $\delta Q\equiv Q-\tau$. The solution of these equations with
boundary conditions $Q(\tau)\to Q_{a,b}(\tau)$ for $\tau\to\mp\infty$
is
\begin{eqnarray}
\label{true_bif}
Q_{\rm opt}(\tau)&=&\tau
-\eta\int\nolimits_0^{\tau}d\tau'[1-\sqrt{2}\,e^{-\tau'^2}]e^{\tau^2-\tau'^2},
\nonumber\\
&&P_{\rm opt}(\tau) = \eta e^{-\tau^2}/\sqrt{2},
\end{eqnarray}
where we took into account the explicit form of $Q_{a,b}(t)$ (\ref{weak_q_ab}).

It is seen from Eq.~(\ref{true_bif}) that the momentum on the optimal
path $P_{\rm opt}$ has a shape of a Gaussian pulse centered at
$\tau=0$, with width $\sim 1$. The coordinate $Q_{\rm opt}(\tau)$ over
the dimensionless time $\tau\sim 1$ switches between the equilibrium
states $Q_{a,b}$.  The typical duration of motion in real time is
$t_l$.

From Eqs.~(\ref{true_bif}), the nonadiabatic activation energy of
escape for $\omega_F\tr \ll 1$ is
\begin{equation}
\label{R_nonad}
\tilde R^{\rm nonad} =(\pi/8)^{1/2}\eta^2
\propto (A_c - A)^2.
\end{equation}
Here, the critical amplitude $A_c$ is given by
Eq.~(\ref{weak_bif}), to first order in $\omega_F$.

\begin{figure}[ht]
\includegraphics[width=3.0in]{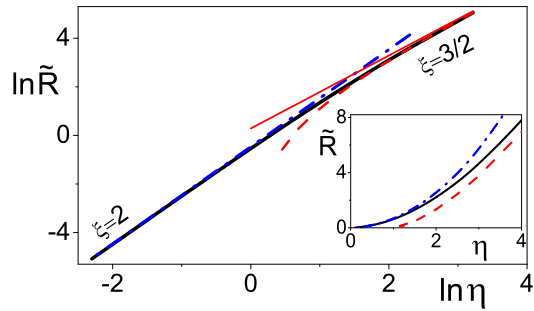}
\caption{The activation energy $\tilde R = -\tilde D\ln\overline W$
on a logarithmic and linear scale (inset) vs. $\eta\propto A_c-A$ for
slow modulation, $\omega_F \tr \ll 1$.  Thick solid lines show the
numerical solution of the variational problem (\protect\ref{R_1d}). It
describes the crossover between different scaling regions.  The thin
solid line shows the adiabatic scaling for large $\eta$,
$\tilde{R}\propto\eta^{3/2}$.  The full result of the adiabatic
approximation is shown by the dashed line.  The dash-dot line shows
the nonadiabatic result (\protect\ref{R_nonad}) that applies for
$\eta\ll 1$; here $\tilde R \propto \eta^{2}$.}
\label{fig:true_bif_states}
\end{figure}

It is seen from Eq.~(\ref{R_nonad}) that, in the locally nonadiabatic
region, the activation energy again displays scaling behavior, $\tilde
R\propto (A_c-A)^{\xi}$. But the scaling exponent is $\xi=2$, it
differs from the adiabatic exponent $\xi=3/2$ (\ref{R_ad}) that has
been known for stationary systems. This is a result of the complicated
nonadiabatic dynamics associated with avoided crossing of the stable
and unstable states. The onset of this scaling behavior is the central
result of the paper.

The predicted $\xi=2$ scaling is compared with the result of the
numerical calculation in Fig.~\ref{fig:true_bif_states}. The analytical
and numerical results are in quantitative agreement in the whole range
$\eta \alt 2$.

\subsection{Activation energy for $\omega_F\tr \gg 1$}

It was shown in Sec.~III C that, sufficiently close to a
bifurcation point, there holds a condition $\omega_F\tr \gg 1$,
even where the modulation frequency is less than the relaxation
rate far away from the bifurcation point,
$\omega_F\tr^{(0)}\lesssim 1$. Finding the activation energy of
escape $\tilde R\equiv \tilde R^{\rm fast}$ for $\omega_F\tr\gg
1$ is formally similar to that in the adiabatic approximation.
The only difference is that $\eta - 1$ in Eq.~(\ref{M_ad}) should
be now replaced by $\eta$. This gives
\begin{equation}
\label{R_fast}
\tilde R^{\rm fast} = (4/3)\eta^{3/2}
\propto (A_c-A)^{3/2}.
\end{equation}
%

Both the coefficient $\beta'$ that relates $\eta$ to $A_c-A$ [see
Eqs.~(\ref{slow_reduced}) and (\ref{coefficients})] and the noise
intensity $\tilde D$ (\ref{noise_fast}) depend on the arbitrary
initial time $t_{\rm i}$. It enters the weighting factor
$\kappa_{11}(t,t_{\rm i})$ which was used in obtaining the equation of
motion for the slow variable (\ref{slow_reduced}). A straightforward
analysis shows that $t_{\rm i}$ drops out from the ratio
$\beta'^{3/2}/\tilde{D}$, which gives the escape rate
$\overline W\propto \exp(-\tilde R^{\rm fast}/\tilde D)$.

Eq.~(\ref{R_fast}) shows that the activation energy displays scaling
behavior with the distance to the bifurcation point in the range
$\omega_F\tr \gg 1$. The scaling exponent is $\xi=3/2$, as in the
adiabatic case.

\subsection{Scaling crossovers near a critical point}

Eqs.~(\ref{R_ad}), (\ref{R_nonad}), and (\ref{R_fast}) show the
onset of three regions where the activation energy of escape
displays scaling dependence on the modulation amplitude,
$R\propto\tilde R\propto (A_c-A)^{\xi}$. The adiabatic and
locally nonadiabatic regions emerge only if the modulation
frequency is slow compared to the relaxation rate far from the
bifurcation point, $\omega_F\tr^{(0)}\ll 1$. In this case, as
seen from Fig.~\ref{fig:true_bif_states}, as the bifurcation
point is approached, the system displays first the adiabatic
scaling $\xi=3/2$, which for smaller $A_c-A$ goes over into the
scaling $\xi=2$. As the bifurcation point $A_c$ is approached
even closer,  there emerges the fast-oscillating regime where
$\xi=3/2$ again.

The widths of the regions of different scaling strongly depend on the
modulation frequency. For small $\omega_F\tr^{(0)} \ll 1$ the range
of amplitudes where motion is effectively fast oscillating,
$\omega_F\tr \gg 1$, is exponentially narrow. However this range
increases very rapidly with increasing $\omega_F$. The particular way
in which the widths of different scaling regions vary with $\omega_F$
depends on the system dynamics. Ultimately, for
$\omega_F\tr^{(0)}\agt 1$, the regime of effectively fast oscillations
becomes the only observable regime near a bifurcation point.

\section{Scaling crossovers for a  model system}

To test the occurrence of three scaling regions and the scaling
crossovers, we have studied activated escape for a model system,
an overdamped Brownian particle in a modulated potential well. It
is described by the Langevin equation
\begin{eqnarray}
\label{eom_model}
&&\dot{q}=-{\partial U(q,t)\over \partial q}+f(t), \quad
\langle f(t)f(t')\rangle=2D\delta(t-t'),
\nonumber\\
&&U(q,t)=-{1\over 3}q^3+ {1\over 4}q - Aq\cos\omega_Ft.
\end{eqnarray}

The shape of the potential $U(q,t)$ is shown schematically in
Fig.~\ref{fig:barrier}. In the absence of modulation, $A=0$, the
system has a metastable state at the bottom of the potential well,
$q_a=-1/2$, and an unstable equilibrium at the barrier top,
$q_b=1/2$. The relaxation time is $\tr^{(0)}=1/U''(q_a)=1$. In the
presence of modulation, the states $q_{a,b}$ oscillate in time. As the
modulation amplitude $A$ increases to the critical value $A_c$ (the
saddle-node bifurcation), the states merge, and then, for $A>A_c$,
disappear.

\begin{figure}[ht]
\includegraphics[width=3.0in]{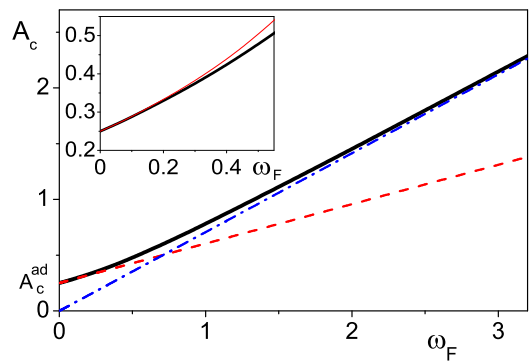}
\caption{The critical amplitude $A_c$ as a function of the modulation
frequency $\omega_F$ for the system (\protect\ref{eom_model}).
Numerical results are shown by thick solid lines. The dashed line
 shows the linear in $\omega_F$ nonadiabatic correction to
$A_c$ described by Eq.~(\protect\ref{weak_bif}). The thin solid
line in the inset describes a correction obtained from the
self-consistent local analysis, Eq.~(\protect\ref{A_c_slow}). The
dash-dot line describes the high-frequency asymptotic that
follows from Eq.~(\ref{real_fast}).} \label{fig:A_c}
\end{figure}

The frequency dependence of the critical amplitude $A_c$ is shown in
Fig.~\ref{fig:A_c}. In the limit $\omega_F=0$ we have $A_c\equiv
A_c^{\rm ad}=1/4$. The linear in $\omega_F$ correction to $A_c$ can be
obtained from Eq.~(\ref{weak_bif}) by noticing that the adiabatic
bifurcational value of the coordinate is $q_c^{\rm ad}=0$, and the
adiabatic bifurcation occurs for $t=0$ (or equivalently,
$t=n\tau_F$). Near the adiabatic bifurcation point we have
\begin{equation}
\label{K_expansion}
\dot q = q^2 + \delta\!A^{\rm ad}-{1\over 2}A\omega_F^2t^2 +f(t),
\end{equation}
with $\delta\!A^{\rm ad}=A-A_c^{\rm ad}$. This equation will have the same
form as Eq.~(\ref{q_1_ad_}) if we replace the factor $A$ in
$A\omega_F^2t^2$ with $A_c^{\rm ad}=1/4$ [as it was done in
Eq.~(\ref{q_1_ad_})].

From Eq.~(\ref{K_expansion}) it follows that, for the model under
consideration, the parameters in Eq.~(\ref{q_1_ad_}) are $\alpha=
\beta = 1, \gamma =(A_c^{\rm ad}/2)^{1/2}=2^{-3/2}$. Therefore, from
Eq.~(\ref{weak_bif}), to first order in $\omega_F$ the critical
amplitude is $A_c^{\rm sl} = 1/4 + 2^{-3/2}\omega_F$. It is shown in
the main part of Fig.~\ref{fig:A_c} with the dashed line.

As discussed in Sec.~III~D, the local nonadiabatic theory allows us to
find higher-order terms in the critical amplitude. This is done by
noticing that the critical state $q_c(t)$ into which the stable and
unstable states merge at the bifurcation point is linear in $t$ for
small $t$, i.e., $q_c(t)-q_c^{\rm ad}\propto t$. By substituting this
solution into Eq.~(\ref{K_expansion}) (without noise) we obtain
\begin{equation}
\label{A_c_slow}
A_c\approx \left[1+\omega_F^2 +
\omega_F\left(\omega_F^2+2\right)^{1/2}\right]/4, \; \omega_F\tr \ll 1.
\end{equation}
This equation is in good agreement with the numerical data for
$\omega_F\tr^{(0)}\alt 0.4$, as seen from the inset in
Fig.~\ref{fig:A_c}. The difference between the numerical and
analytical $A_c$ decreases exponentially fast with decreasing $\omega_F$.

We also evaluated for slow driving the time derivative of the
eigenvalue $\mu_1=2q_c(t)$ on the critical cycle $q_c(t)$. For the
model (\ref{K_expansion}) the stable and unstable states are at their
closest for $t=0$.  Eqs.~(\ref{eta_def}), (\ref{xi_renormalized}) show
that, at this time, $\dot\mu_1=(2A_c)^{1/2}\omega_F$. This value
agrees with the numerical values of $\dot\mu_1$ to better than 10\%
for $\omega_F < 0.5$.

In the high frequency limit, $\omega_F\tr^{(0)} \gg 1$, the motion of
the system (\ref{eom_model}) is a superposition of slow motion and
fast oscillations at frequency $\omega_F$. To lowest order in
$\omega_F^{-1}$ we have $q\approx Q+(A/\omega_F)\sin\omega_Ft$. The
equation for the slow variable $Q$ becomes
\begin{equation}
\label{real_fast}
\dot Q = Q^2-{1\over 4} +
{A^2\over 2\omega_F^2} + f(t).
\end{equation}
It shows that, for large $\omega_F$, we have $A_c\approx
\omega_F/\sqrt{2}$. This is in good agreement with
numerical data in Fig.~\ref{fig:A_c} for $\omega_F\agt 2$.

In the intermediate range, $\omega_F\tr^{(0)}\alt 1$, the motion may
not be separated into slow and fast-oscillating for weak modulation,
but the separation becomes possible near a critical point,
$\omega_F\tr \gg 1$. Here, the coefficients in the equation of motion
for the slow variable and the effective noise intensity
(\ref{coefficients}), (\ref{noise_fast}) are nonlocal and had to be
evaluated numerically as functions of $\omega_F$.

\subsection{Activation energy}

For a periodically modulated overdamped Brownian particle described by
Eq.~(\ref{eom_model}), the activation energy of escape $R$ can be
found from the variational problem (\ref{var_problem}),
(\ref{escape_rate}) or, equivalently, (\ref{R_1d})
\cite{LS_chaos}. The variables $Q,P, \tau$, and the function $G$ in
the Lagrangian $L$ (\ref{R_1d}) and the Hamilton equations
(\ref{eqn_H}) should be changed to $q,p, t$, and $-\partial
U(q,t)/\partial q$, respectively. As explained in Sec.~II, there is
one optimal path per modulation period. The initial condition for the
momentum $p$ on the optimal path is given by Eq.~(\ref{initial}), with
$Q_a(\tau)$ replaced by $q_a(t)$ [the expression for $\sigma_a^2$ can
be further simplified taking into account the periodicity of
$q_a(t)$].  Then Eqs.~(\ref{eqn_H}) can be solved numerically.

\begin{widetext}

\begin{figure}
\includegraphics[width=6.5in]{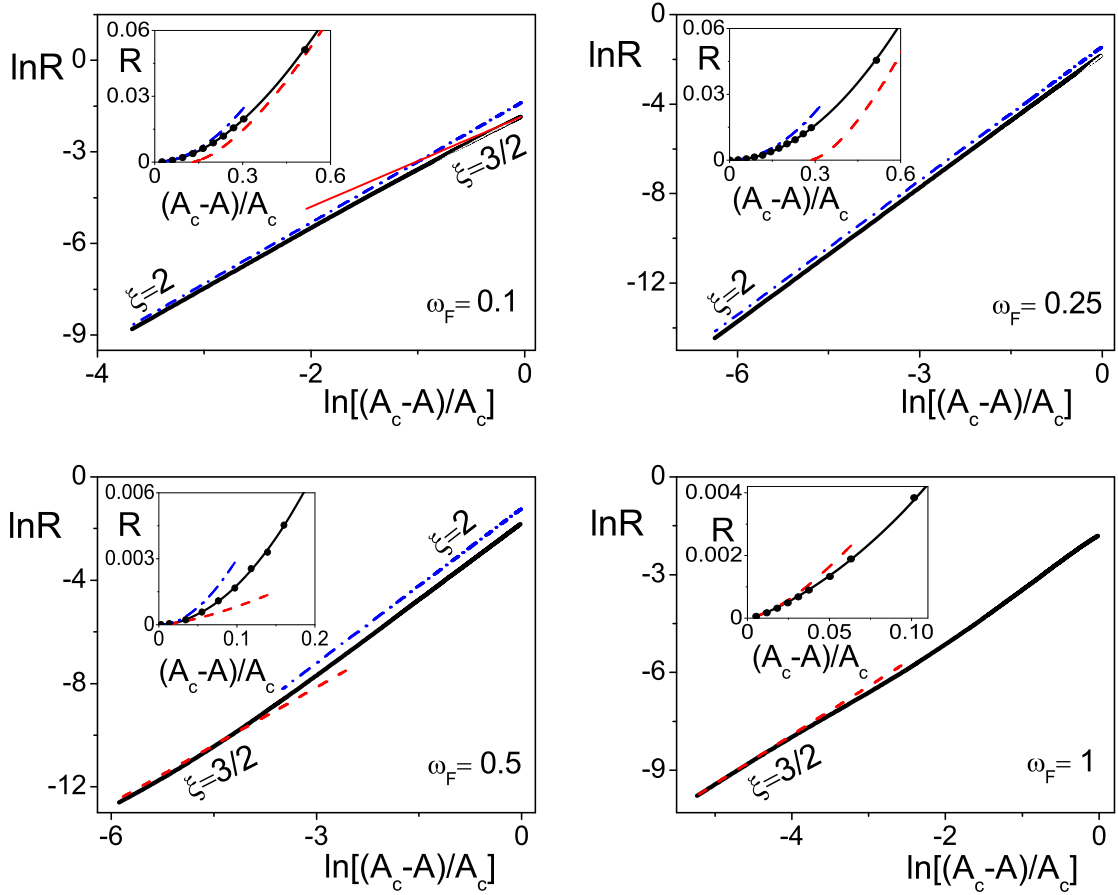}
\caption{The activation energy of escape $ R$ vs. modulation amplitude
$A$ on the logarithmic and linear (inset) scales for a Brownian
particle in a modulated potential (\protect\ref{eom_model}). The
values of $\omega_F$ are indicated on each panel. The thick solid
lines show the results of the numerical solution of the variational
problem for $R$. The dashed lines for $\omega_F=0.1, 0.25$ show the
adiabatic approximation, whereas for $\omega_F=0.5,1.0$ they show the
approximation of effectively fast oscillations: in the both cases the
scaling exponent is $\xi=3/2$ (for $\omega_F=0.1$ this asymptotic
scaling is shown by the thin solid line). The dash-dot lines show the
$\xi=2$ scaling (\protect\ref{R_nonad}). The dots show the results of
numerical simulations of Eq.~(\protect\ref{eom_model}). }
\label{fig:panels}
\end{figure}

\end{widetext}

The obtained activation energy $R$ as function of the modulation
amplitude for four characteristic values of $\omega_F$ is shown in
Fig.~\ref{fig:panels}. The solid lines on this plot correspond to the
results of the numerical solution of the variational problem.
The dashed lines in the panels for $\omega_F=0.1, 0.25$ (we remind
that $\tr^{(0)}=1$) show the adiabatic approximation,
\[R^{\rm
ad}=\min\left[U\bigl(q_b(t)\bigr)-U\bigl(q_a(t)\bigr)\right] = {4\over
3}\left({1\over 4}-A\right)^{3/2}.\]

The dash-dot lines on all panels show the locally nonadiabatic
approximation near the bifurcation point, which gives
\[R^{\rm nonad}= \left(\frac{\pi}{8}\right)^{1/2}\left(\frac{2}{\dot\mu_1}\right)^{1/2}(A_c-A)^2.\]
In plotting this expression we used the values of $A_c$ and
$\dot\mu_1$ found numerically (they were very close to the analytical
expressions given above).

Finally, the dashed lines in Fig.~\ref{fig:panels} in the panels for
$\omega_F=0.5,1$ show scaling for the effectively fast-oscillating
regime near the bifurcation point, with
\[R^{\rm fast}= {4\over 3}\beta'^{3/2}{D\over\tilde D}\,
\left(A_c-A\right)^{3/2}.\] The coefficients $\beta'$ and $\tilde D$ as
given by Eqs.~(\ref{coefficients}), (\ref{noise_fast}) were obtained
numerically.

It is seen from Fig.~\ref{fig:panels} that, for small $\omega_F$, the
adiabatic approximation applies over a broad region of driving
amplitudes. Near the bifurcation point it gives scaling $R\propto
(A_c-A)^{\xi}$ with $\xi=3/2$ (cf. the panel for
$\omega_F=0.1$). However, close to the bifurcation point this scaling
does not work. Instead there emerges the nonadiabatic dynamic scaling
with $\xi=2$. For $\omega_F=0.1$ the range of the
nonadiabatic scaling is comparatively narrow.

As the frequency increases, the amplitude range characterized by the
$\xi=2$-scaling dramatically increases. For $\omega_F=0.25$ this is
practically the only scaling seen near the bifurcation point.

With further increase of $\omega_F$, close to the bifurcation point
there emerges a region of the fast-oscillation scaling $R\propto
(A_c-A)^{\xi}$ where again $\xi=3/2$. The panel for $\omega_F=0.5$
shows a crossover from the scaling $\xi=3/2$ very close to the
bifurcation point to the scaling $\xi=2$ further away from $A_c$. Note
that the frequency $\omega_F=0.5$ is neither small nor large, and
therefore there is a noticeable difference in the coefficients at
$(A_c-A)^{2}$ obtained from the full variational problem for $R$ and
from the locally nonadiabatic theory near the bifurcation point.

When $\omega_F\tr^{0)}\agt 1$ we do not expect to see scaling for
either adiabatic or locally nonadiabatic regime. The only scaling to
be expected near the bifurcation point is the fast-oscillation one,
with $\xi=3/2$. This is indeed seen in the panel for
$\omega_F=1$ in Fig.~\ref{fig:panels}. We note that the
global fast-oscillation approximation (\ref{real_fast}) does not apply
for $\omega_F\tr^{0)}= 1$.

\subsection{Simulations}

An additional test of the results can be obtained by directly
simulating the Brownian dynamics described by
Eq.~(\ref{eom_model}). We conducted such simulations using the second
order integration scheme of stochastic differential equations
developed in Ref.~\onlinecite{Mannella_algorithm}. As a result of the
simulations we obtained the probability distribution of the dwell time
in the metastable state $p_{\rm dw}(t)$. It gives the probability
density (over time) for a system prepared at $t=0$ close to the
attractor to stay in the basin of attraction until time $t$ and leave
at that time.

In practice we calculated $p_{\rm dw}(t)$ by detecting the system at
time $t$ at a point $q$ that lied well beyond, but not too far from,
the oscillating boundary $q_b(t)$. It is simply related to the
time-dependent escape probability $W(t)$, which gives the probability
current from the attraction basin at time $t$ if the system was in the
stable state at $t=0$ \cite{prefactor},
\begin{equation}
\label{dwell}
p_{\rm dw}(t)= W(t)\exp\left[-\int\nolimits_0^t dt'\,W(t')\right].
\end{equation}
The average escape rate is given by the mean dwell time,
\begin{equation}
\label{average_rate}
\overline W = \left[\int\nolimits_0^{\infty}dt\,
t \,p_{\rm dw}(t)\right]^{-1}.
\end{equation}
We studied small noise intensities so that $\overline W\ll \omega_F$;
then $\overline W$ was independent of the position of the
``observation'' point $q$.

\begin{figure}
\includegraphics[width=2.5in]{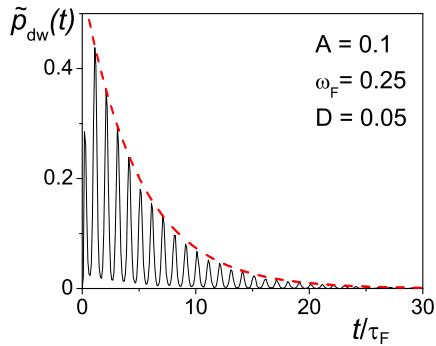}
\caption{The scaled probability density of the dwell time $\tilde
p_{\rm dw}(t)=p_{\rm dw}(t)\tau_F$ obtained by numerical simulations
of a Brownian particle in a modulated potential,
Eq.~(\protect\ref{eom_model}). The parameters are $A=0.1, D=0.05,
\omega_F=0.25$ (solid line). The dashed line shows the exponential fit
of the envelope with decrement $\overline W \tau_F=0.008$.}
\label{fig:mod}
\end{figure}

In most simulations the system was prepared initially at the stable
state $q_a(t)$; we found that $\overline W$ was
independent of the initial state provided it was close to $q_a(t)$.

The dwell-time distribution for a particular set of parameter values
in Eq.~(\ref{eom_model}) is shown in Fig.~\ref{fig:mod}. The data
refer to modulation at a comparatively low frequency and with
comparatively small amplitude. The function $p_{\rm dw}(t)$ is
strongly modulated in time, with period $\tau_F$. This means that
escape events are strongly synchronized by the modulation, in
agreement with the analytical results for $W(t)$ obtained for the same
model in Ref.~\onlinecite{prefactor}.

The average escape rate $\overline W$ was found from
the data of the type shown in Fig.~\ref{fig:mod} by calculating the
mean dwell time (\ref{average_rate}) and also from an exponential fit
of the envelope of $p_{\rm dw}(t)$. These two approaches gave the same
result. For each set of $A,\omega_F, D$ we observed $\sim 10^5$ escape
events. Then $D$ was changed. The activation energy of escape was
found from $\ln\overline W$ for 2-4 values of $D$. We
tested that it was independent of $D$ in the range $R/D\agt 6$.

The data of simulations are shown in Fig.~\ref{fig:panels} by
dots. For all parameter values they are in excellent agreement with
the results of the numerical solution of the variational problem
(\ref{var_problem}).

\section{Conclusions}

In conclusion, we have identified three regions near a bifurcation
point where the activation energy of escape displays scaling behavior
as a function of the amplitude of periodic modulation.  The main
results refer to slow modulation, where $\omega_F\tr\ll 1$.  We show
the emergence of nonadiabatic behavior in this region. The
nonadiabaticity leads to a crossover from the scaling with exponent
$\xi=3/2$, previously found for stationary systems, to a new dynamical
scaling with $\xi=2$. The $\xi=2$ region first emerges near the
bifurcation point and then expands with increasing modulation
frequency. With further increase of $\omega_F$ the crossover $\xi=2$
to $\xi=3/2$ can be observed. Again, the effectively fast-oscillating
region with $\xi=3/2$-scaling emerges first near the bifurcation
point. Even though the widths of the regions of different scaling
depend on the parameters of a system, the phenomenon of scaling
crossovers should be universal.

The onset of the $\xi=2$-scaling is a consequence of the slowing down
of motion near a bifurcation point. The adiabatic relaxation time of
the system $\tr$ strongly depends on the distance to the bifurcation
point, $\tr\propto (A_c-A)^{-1/2}$. The nonadiabatic scaling emerges
where $\tr$ becomes $\sim [\tau_F\tr^{(0)}]^{1/2}$. For smaller
$|A_c-A|$ the dependence of $\tr$ on $A_c-A$ becomes weak, while $\tr$
still largely exceeds the modulation period $\tau_F$. This is
associated with avoided crossing of the stable and unstable states,
which occurs with decreasing $A_c-A$ as these states are pressed
against each other when the system approaches the adiabatic
bifurcation point, see Fig.~\ref{fig:nonad}.

Both in the adiabatic regime and the nonadiabatic regime for
$\omega_F\tr \ll 1$ escape is most likely to occur while the
equilibrium states are close to each other. The escape rate is
therefore determined by the behavior of the system for a small portion
of the modulation period, i.e., locally in time.

The regime of effectively fast oscillations near the bifurcation point
emerges for $\omega_F\tr \gg 1$. It can arise even where the
modulation period $\tau_F$ exceeds the relaxation time far from the
bifurcation point. In this regime the motion is controlled by a slow
variable, but the dynamics of this variable is no longer determined by
local (in time) characteristics. The relevant characteristics are
obtained by averaging the appropriate parameters along the critical
periodic trajectory of the system into which the stable and unstable
periodic cycles merge at the bifurcation point.

We have developed a general formulation of escape of periodically
modulated systems driven by colored Gaussian noise.  Near a
bifurcation point slow motion of the system effectively filters out
high-frequency components of the noise spectrum and makes the noise
effectively white. From the theoretical point of view it is
interesting that, in the locally-nonadiabatic regime of $\xi=2$
scaling, the instanton-like optimal escape path can be found from {\it
linear} equations of motion.

We expect that the new $\xi=2$ scaling and the scaling crossovers can
be seen in various systems. Examples are modulated Josephson
junctions, nanomagnets, and optically trapped Brownian particles,
where escape in the presence of modulation has been already studied
experimentally, albeit in different regimes.

This research was supported by the NSF
PHY-0071059 and NSF DMR-0305746.

\appendix


\section{Variational equations for the escape problem}

We will consider here the optimal trajectory that the system is most
likely to follow in escape, ${\bf q}_{\rm opt}(t)$, and the random
force ${\bf f}_{\rm opt}(t)$ that drives the system during this
motion. The trajectories ${\bf q}_{\rm opt}(t), {\bf f}_{\rm opt}(t)$
provide the absolute minimum to the functional ${\cal R}[{\bf q,f}]$
(\ref{var_problem}). The variational equations have the form
\begin{eqnarray}
\label{var_eqs1}
&& \int dt'\hat{\cal F}(t-t')\,{\bf f}(t')-{\bm \lambda}(t)=0,\\
\label{var_eqs2}
&&\dot{\bm \lambda}= -{\bm\nabla}({\bm \lambda} \cdot {\bf K}),\\
\label{var_eqs3}
&&\dot{\bf q}- {\bf K}({\bf q};A,t)-{\bf f}(t)=0
\end{eqnarray}
(here and below we use the hat to indicate a matrix; ${\bm\nabla}\equiv
\partial/\partial{\bf q}$).

From Eqs.~(\ref{phi_inverse}) and (\ref{var_eqs1}) it follows that the
optimal noise realization is expressed in terms of the Lagrange
multiplier ${\bm \lambda}$ and the matrix of the noise correlation
functions $\hat\varphi$ (\ref{corr_matrix}) as
\begin{equation}
\label{f_opt}
{\bf f}_{\rm opt}= D^{-1}\int dt'\hat\varphi(t-t')\,{\bm \lambda}(t').
\end{equation}
Therefore, for the optimal path, the activation energy functional
(\ref{R_functional}) is
\begin{equation}
\label{lambda_functional}
{\cal R} = \frac{1}{2}D^{-1}\int\!\!\!\int dtdt'
{\bm \lambda}(t)\cdot\,\hat\varphi(t-t'){\bm \lambda}(t')
\end{equation}
(note that the noise intensity $D$ drops out from ${\cal R}$, because
the correlation functions $\varphi_{ij}(t)$ are proportional to the
noise intensity themselves).

From the structure of the functional (\ref{lambda_functional})
(integration over time goes from $-\infty$ to $\infty$ and the
integral is nonnegative) and from the fact that the system is
initially in the vicinity of a stable state one immediately derives
the boundary condition (\ref{boundary}) for $t\to -\infty$. The
arguments generalize to periodically driven systems the analysis of
Ref.~\onlinecite{Dykman-90} for stationary systems.

Close to a periodic stable state ${\bf q}_a(t)$ time evolution of
${\bm \lambda}(t)$ can be described using the matrix $\hat
\mu_a(t)=\bigl(\partial K_i/\partial q_j\bigr)_a$, where the
derivatives are evaluated for the state ${\bf q}_a(t)$. Because of the
periodicity of ${\bf q}_a(t)$, the matrix $\hat\mu_a(t)$ is also
periodic in time.  It determines the monodromy matrix
\[\hat M_a(t)=T_t\exp\left[\int_t^{t+\tau_F}dt_1\,\hat\mu_a(t_1)\right],\]
where $T_t$ is the operator of chronological ordering (cf. Appendix
C). The matrix $\hat M_a$ shows how the distance between a point ${\bf
q}$ and the cycle ${\bf q}_a(t)$ varies over the modulation period in
the absence of noise for small $|{\bf q} -{\bf q}_a(t)|$:
\[{\bf q}(t+\tau_F)-{\bf
q}_a(t) = \hat M_a(t)[{\bf q}(t)-{\bf q}_a(t)]\]
From the condition that
${\bf q}_a(t)$ is a stable state, the eigenvalues of the matrix $\hat
M_a$ are all less then 1 in absolute value: in this case the distance
between ${\bf q}$ and ${\bf q}_a(t)$ decreases with increasing time.

It is seen from Eq.~(\ref{var_eqs2}) that the monodromy matrix for
${\bm \lambda}$ is the inverse transposed of $\hat M_a$. Therefore its
eigenvalues are all larger than 1 in absolute value. Hence, if
 the system is in the stable state ${\bf q}_a(t)$ for $t\to -\infty$,
 then ${\bm \lambda}(t)\to {\bf
0}$ for $t\to -\infty$, and from (\ref{var_eqs1}) ${\bf f}(t)\to {\bf
0}$, too.

For the periodic saddle ${\bf q}_b(t)$ on the boundary of the
attraction basin, one of the eigenvalues of the corresponding
monodromy matrix exceeds $1$ in absolute value. It is such a
saddle-type boundary that can merge with an attractor at a saddle-node
bifurcation we are interested in. Respectively, one eigenvalue of the
matrix that describes time evolution of ${\bm \lambda}$ is
$<1$. If ${\bm\lambda}$ is pointing along the corresponding
eigenvector, it will decay as $t\to \infty$. Then ${\bf f}(t)$ will
decay, too, from Eq.~(\ref{var_eqs1}). This means that the system may
asymptotically approach a saddle-type state. Note that there are no
optimal paths that would go from one stable state to another, because
the condition ${\bm \lambda}\to {\bf 0}$ for $t\to \infty$ is not
satisfied there. This explains the boundary condition (\ref{boundary})
for $t\to\infty$.

Because the function ${\bf K}$ is periodic in time,
Eqs.~(\ref{var_eqs1})--(\ref{var_eqs3}) with boundary conditions
(\ref{boundary}) have a periodic set of solutions. If ${\bf q}(t),
{\bf f}(t), {\bm \lambda}(t)$ is a solution, then ${\bf q}(t+\tau_F),
{\bf f}(t+\tau_F), {\bm \lambda}(t+\tau_F)$ is a solution, too.  These
solutions are heteroclinic orbits: they connect the states ${\bf
q}_a(t), {\bf f}= {\bm \lambda}= {\bf 0}$ and ${\bf q}_b(t), {\bf f}=
{\bm \lambda}= {\bf 0}$, which are also solutions of
Eqs.~(\ref{var_eqs1})--(\ref{var_eqs3}). Generally, only one
heteroclinic orbit per period provides the minimum to the functional
${\cal R}$ (\ref{var_problem}).

\subsection{Escape in systems with a slow variable}

Eqs.~(\ref{var_eqs1})--(\ref{var_eqs3}) are largely simplified if one
of the motions in the system is slow and all other variables follow
this motion adiabatically, i.e., their relaxation time $\tr^{(0)}$ is
much smaller than the relaxation time of the slow variable $\tr$, at
least for a part of the modulation period near a bifurcation point. We
will assume that slow motion is described by $q_1$; the variable $q_1$
itself may be a periodic function modulated by a slowly varying factor
$Q_1$, as in the case discussed in Sec.~III~C. In this case we will be
interested primarily in the factor $Q_1$.

Over a time $\sim\tr^{(0)}$, the variables $q_{i>1}$ reach their
equilibrium values $q'_i(q_1,t)$ for given $Q_1,t$. They are
determined by the equations
\begin{equation}
\label{q_fast}
\dot q'_i=K_i(q_1,q'_{i>1}; A,t)+ f_i \quad (i>1)
\end{equation}
In the absence of noise, $f_i=0$, the solutions of these equations
 are periodic for given $Q_1$. As we will see below,
 the terms $f_i$ here are small and give small corrections.

It is important that the periodic solutions $q'_{i>1}$ with
$f_{i>1}=0$ be stable. From this condition and Eq.~(\ref{var_eqs2}) it
follows immediately that $\lambda_{i>1}=0$, otherwise the components
$\lambda_{i>1}$ would exponentially grow in time, leading to the onset
of a large force that would drive the system away from the state
(\ref{q_fast}) with given $Q_1$.

The nonzero component of the Lagrange multiplier $\lambda_1$ is
determined by the component of the optimal force $f_1$. The latter
should overcome the restoring force on the slow variable $Q_1$ and
drive it from the stable to the unstable state. But the slowness of
$Q_1$ means that the restoring force is small, and therefore the force
$f_1$ should be small, too. It is smaller than the force that would be
needed to overcome the restoring force for the fast variables by at
least a factor $\sim \tr^{(0)}/\tr$. Therefore $\lambda_1$ is
small as well.

For given $\lambda_1(t)$ we can find all components $f_{i>1}$ from
Eq.~(\ref{f_opt}). They are all $\propto 1/\tr$, and therefore to
lowest order in $\tr^{(0)}/\tr$ they can be disregarded in the
solution of Eq.~(\ref{q_fast}) for $q_{i>1}$.

The problem of escape is therefore reduced to a one-dimensional
problem for the variable $q_1$, the force $f_1$, and the Lagrange
multiplier $\lambda_1$. In Eq.~(\ref{var_eqs3}) for $q_1$, the
functions $q_{i>1}$ should be replaced by $q'_{i>1}$ calculated for $f_i=0$.

Further simplification occurs if the noise spectrum is smooth.  The
analysis here is different for the cases of slow and fast modulation,
i.e. whether $\omega_F\tr$ is small or large. The case $\omega_F\tr\gg
1$ is discussed in Appendix~C.  Here and in Appendix B we consider the
case $\omega_F\tr\ll 1$.  The major effect of noise on the slow
variable $q_1$ comes from the noise spectral components at frequencies
$\omega \lesssim 1/\tr$. If the noise spectrum is flat for such
frequencies, the noise can be assumed to be white on the ''slow'' time
scale. In other words, the correlation function $\varphi_{11}(t)$ can
be replaced by $2D\delta (t)$ [in this situation it is convenient to
choose $D$ from this condition rather than to define it by
Eq.~(\ref{noise_intensity})]. Then ${\cal F}_{11}(t)= {1\over
2}\delta(t)$.

For a 1D system driven by white noise of intensity $D$,
Eqs.~(\ref{var_eqs1})--(\ref{var_eqs3}) have a solution
\begin{eqnarray}
\label{var_1D}
&&f_1(t)=2\lambda_1(t) = \dot q_1-K_1,\\
&&{\cal R}={1\over 4}\int dt\,\left(\dot q_1-K_1\right)^2\nonumber
\end{eqnarray}
This
reduces the variational problem of finding the optimal path to the
known formulation for white-noise driven systems \cite{Freidlin84}.


\section{Reduced equation of motion for slow driving}

In this Appendix we derive simplified equations of motion for the case
of slow driving where the relaxation time $\tr\ll\tau_F$ and the
motion can be described in the adiabatic approximation. We will
consider the vicinity of the adiabatic bifurcation point ${\bf q}={\bf
0}, t=0, A=A_c^{\rm ad}$. A convenient basis for ${\bf q}$ is provided
by the set of the right eigenvectors of the matrix
$\hat{\mu}=(\partial K_i/\partial q_j)$, where the derivatives are
evaluated at the adiabatic bifurcation point. In this basis the
equation of motion (\ref{eom}) has the form
\begin{eqnarray}
\label{expansion}
\dot q_i&\approx&\mu_iq_i +{1\over 2}\sum\nolimits_{j,k}K_{i;jk}q_jq_k
+ K_{i;A}\delta\!A^{\rm ad} + K_{i;t}t \nonumber\\
&+&{1\over 2}K_{i;tt}t^2 + \sum\nolimits_jK_{i;jt}q_jt + f_i(t).
\end{eqnarray}
Here $K_{i;jk}=\partial^2K_i/\partial q_j\partial q_k$,
$K_{i;A}= \partial K_i/\partial A$, etc, with all derivatives
calculated at the adiabatic bifurcation point, and $\delta\!A^{\rm ad}=
A - A_c^{\rm ad}$. Since the function ${\bf K}$ depends on $t$ only in
terms of the modulation phase $\phi=\omega_Ft$, we have
$K_{i;t}\propto\omega_F$. The expansion in $t$ in (\ref{expansion}) is,
in fact, an expansion in $\omega_Ft$.

Because the eigenvalue $\mu_1$ is equal to zero, for small $|{\bf q}|$
relaxation of $q_1$ is much slower than relaxation of $q_{i>1}$. In
the absence of noise, over the relaxation time $\tr^{(0)}$ the
variables $q_{i>1}$ approach their quasistationary values for given
$q_1$ and $\phi=\omega_Ft$. They can be obtained from the equations of
motion (\ref{expansion}) for $q_{i>1}$ in which $\dot q_i$ and
$f_i(t)$ are disregarded,
\begin{equation}
\label{q_{i>1}}
q_i \approx -\mu_i^{-1}\left[K_{i;t}t +K_{i;11}q_1^2 + K_{i;A}\delta\!A^{\rm ad} + \ldots \right]\quad (i>1).
\end{equation}
Here, the major term is linear in $\omega_Ft$. The full expression is
a series in $q_1, \omega_Ft$, and $\delta\!A$, and the omitted terms
are of higher order in these variables.  Because of the noise,
$q_{i>1}$ will additionally perform small fluctuations with amplitude
$\propto D^{1/2}$.

The dynamics of the slow variable $q_1$ on times exceeding $\tr^{(0)}$
is given by Eq.~(\ref{expansion}) with $i=1$, in which $q_{i>1}$ are
replaced with their quasistationary values. To lowest order in $q_1$,
$\delta\!A^{\rm ad}$, and $\omega_Ft$ only the linear in $t$ term
should be kept in Eq.~(\ref{q_{i>1}}).  This gives
\begin{eqnarray}
\label{q_1_ad}
&\dot q_1 = \alpha q_1^2 + \beta\,\delta\!A^{\rm ad} -
\alpha\gamma^2 (\omega_Ft)^2 + f_1(t), \; \\
&\alpha = {1\over 2}K_{1;11},\;\beta =
K_{1;A},\nonumber
\end{eqnarray}
with
\begin{eqnarray}
\label{gamma^2}
\gamma^2&=& - K_{1;11}^{-1}\omega_F^{-2}
\left( K_{1;tt} +\sum\nolimits_{i,j >
1}\mu_i^{-1}\mu_j^{-1}K_{1;ij}K_{i;t}K_{j;t} \right.  \nonumber\\
&&\left.  - 2\sum\nolimits_{j>1}\mu_j^{-1}K_{1;jt}K_{j;t}\right ).
\end{eqnarray}
Note that the coefficient $\gamma$ is independent of $\omega_F$.

Eq.~(\ref{q_1_ad}) reduces the multi-dimensional problem of random
motion near a bifurcation point to a one-dimensional problem. In the
case of a one-dimensional overdamped system driven by an additive
periodic force it was derived earlier in
Refs. \onlinecite{Berglund02,univ_arch02}. Besides the noise term,
it differs from the equation of motion for stationary systems in the
vicinity of a saddle-node bifurcation point \cite{Guckenheimer} in
that it has a term $\propto (\omega_Ft)^2$.

Depending on
the sign of $\delta\!A^{\rm ad}$, Eq.~(\ref{q_1_ad}) has either two
adiabatic  solutions
\begin{eqnarray}
\label{ad_states}
(q_1)_{a,b}^{\rm ad} = \mp\sgn(\alpha)[-(\beta/\alpha)\delta\!A^{\rm
ad} +(\gamma\omega_Ft)^2]^{1/2}
\end{eqnarray}
or none. For concreteness, we assume that the adiabatic solutions
exist for $\delta\!A^{\rm ad} < 0$, i.e. $\alpha\beta > 0$. The
solutions are even functions of time. They touch each other at $t=0$
for $\delta\!A^{\rm ad} = 0$. We assume that the periodic
adiabatic states ${\bf q}^{\rm ad}_{a,b}(t)$ exist for all times
provided $\delta\!A^{\rm ad} < 0$.

The term $K_{1;t}t$ in Eq.~(\ref{expansion}) has to be equal to zero,
otherwise the bifurcation point will be far from $\delta\!A^{\rm
ad}=t=0$.  On the other hand, the equation of motion (\ref{q_1_ad})
may contain the term $Cq_1\omega_Ft$, where $C$ is a sum of $K_{1;1t}$
and appropriately weighted products $K_{1;1i}K_{i;t}$. This term can
be eliminated by a linear transformation $q_1 \to q_1 +
C\omega_Ft/2\alpha$ and renormalizations $\delta\!A^{\rm ad} \to
\delta A^{\rm ad} + C\omega_F/2\alpha\beta$, $\gamma^2 \to
\gamma^2+(C/2\alpha)^2$. The renormalized $\gamma^2$ should be
positive, if the stable and unstable adiabatic periodic states touch
each other only for $\delta\!A^{\rm ad}=t=0$ and only once per period.

The term $\propto q_1\omega_Ft$ does not arise in the important case
where the modulation is performed by an additive periodic force ${\bf
F}(t)$, see Eq.~(\ref{additive_force}). Here, the adiabatic states
${\bf q}_{a,b}^{\rm ad}(t)$ correspond to the minimum and maximum of
the potential $U_0({\bf q})-{\bf F}(t)\cdot{\bf q}$,
cf. Fig.~\ref{fig:barrier}. They merge first with increasing
modulation amplitude $A$ when the field component $|F_1|$ is at its
maximum over $t$.  This means that $\partial_t{\bf K}=
\partial_{\bf q}\partial_t{\bf K} ={\bf 0}$ at the bifurcation point.

As explained in Sec.~III, the typical relaxation time near the
bifurcation point does not exceed
$t_l=(\alpha\gamma\omega_F)^{-1/2}$. If the correlation time of the
noise $f_1(t)$ is much less than $t_l, \tr^{\rm ad}$ and the power
spectrum of the noise does not have singular features for high
frequencies, then the dimensionless noise
\begin{equation}
\label{f_slow}
\tilde f(\tau)\equiv (\gamma\omega_F)^{-1}f_1(t) \quad (\tau = t/t_l)
\end{equation}
is effectively $\delta$-correlated as a function of the ``slow'' time
$\tau$, with $\langle \tilde f(\tau)\tilde f(0)\rangle = 2\tilde
D\delta(\tau)$. From Eqs.~(\ref{corr_matrix}), (\ref{f_slow}) the
effective noise intensity is
\begin{equation}
\label{tilde_D}
\tilde
D=|\alpha/4|^{1/2}(\gamma\omega_F)^{-3/2}\int\nolimits_{-\infty}^{\infty}
dt\,\varphi_{11}(t).
\end{equation}


\section{Reduced equation of motion for fast driving}

In this Section we consider the case where modulation near the
bifurcation point is effectively fast, so that $\omega_F \tr \gg
1$. Here, throughout the modulation cycle the stable and unstable
states ${\bf q}_{a,b}(t)$ stay close to each other and to the
critical cycle ${\bf q}_c(t)$ into which they merge at the
bifurcation point $A=A_c$. Therefore the equation of motion
(\ref{eom}) can be expanded in $\delta {\bf q}={\bf q}-{\bf
q}_c(t)$, $A-A_c$, leading to Eq.~(\ref{near_bifurcation}). The
expansion coefficients are periodic in time.

It is convenient to start the analysis by simplifying the part of
Eq.~(\ref{near_bifurcation})
\[\delta \dot{\bf q} = \hat\mu\,\delta
{\bf q}, \quad \hat\mu = \hat \mu(t+\tau_F),\]
that describes motion in the linear approximation in $\delta {\bf
q}$. We introduce the matrix $\hat\kappa(t,t_{\rm i})$ such that
\begin{equation}
\label{kappa}
\hat\kappa(t,t_{\rm i})=T_t\exp\left(\int\nolimits_{t_{\rm i}}^tdt_1\hat\mu(t_1)\right),
\end{equation}
where $T_t$ is the operator of chronological ordering. This matrix
satisfies the equation $\partial\hat\kappa(t,t_{\rm i})/\partial t =
\hat\mu(t)\hat\kappa(t,t_{\rm i})$ and gives the monodromy matrix $\hat M$,
\[\hat M(t) \equiv \hat M(t+\tau_F)= \hat\kappa(t+\tau_F,t).\]

The eigenvalues $M_{\nu}$ of the matrix $\hat M$ determine the
evolution of $\delta {\bf q}(t)$ in linear approximation. Over the
period $\tau_F$, the coefficients of the expansion of $\delta {\bf
q}(t)$ in the right eigenvectors ${\bf e}_{\nu}(t)$ of $\hat M$ change
in $M_{\nu}$ times (we use Greek letters to enumerate eigenvalues and
eigenvectors; they should be distinguished from the vector components,
like $q_i$). The eigenvalues $M_{\nu}$ are independent of time because
of periodicity of $\hat M(t), \hat\mu(t)$. They are simply related to
the Floquet exponents for the periodic state ${\bf q}_c(t)$.

At the saddle-node bifurcation, where stable and saddle-type states
coalesce, one of the eigenvalues (for example, $M_1$) becomes equal to
1, whereas $|M_{\nu>1}|< 1$. This means that the system is attracted
to ${\bf q}_c(t)$ in all directions except for the critical direction
${\bf e}_1(t)$; the distance from ${\bf q}_c(t)$ along ${\bf e}_1(t)$ does
not change over the period, in linear approximation. In what follows
we choose ${\bf e}_1$ to be real.

For small $\delta\!A$ of an appropriate sign, the state ${\bf q}_c(t)$
splits along ${\bf e}_1(t)$ into a stable and an unstable state. The
system approaches the vicinity of these states along the directions
${\bf e}_{\nu>1}$ over a short time
$\tau_F\max\left[1/\bigl\vert\ln|M_{\nu>1}|\bigr\vert\right]\sim \tr^{(0)}$. In contrast, the motion
along ${\bf e}_1$ is slow.

The ${\bf e}_1$-component of $\delta {\bf q}$ is the soft mode.
We are interested in its dynamics on times long compared to
$\tr^{(0)}, \tau_F$. The analysis  is simplified by the fact
that, for $t-t_{\rm i}\gg \bigl\vert\tau_F/\ln|M_{\nu >
1}|\bigr\vert$, the matrix $\hat\kappa(t,t_{\rm i})$ projects any
vector $\delta {\bf q}(t_{\rm i})$ on the vector ${\bf e}_1(t)$.
In particular,
\begin{eqnarray}
\label{projection}
\hat\kappa(t,t_{\rm i}){\bf e}_1(t_{\rm i}) \approx \kappa_{11}(t,t_{\rm i}){\bf
e}_1(t).
\end{eqnarray}
This is a consequence of the transitive property $\hat\kappa(t,t_{\rm i}) =
\hat\kappa(t,t')\hat\kappa(t',t_{\rm i})$ and the fact that, for an
arbitrary vector $\delta{\bf q}$, we have $\hat M^n(t)\,\delta{\bf
q} \to C{\bf e}_1(t)$ for $n\to\infty$. The function
$\kappa_{11}$ in Eq.~(\ref{projection}) is given by the expression
\[\kappa_{11}(t,t_{\rm i})= \bar{\bf e}_1(t)\cdot\hat\kappa(t,t_{\rm i}){\bf e}_1(t_{\rm i}).\]
Here, $\bar{\bf e}_1$ is the left eigenvector of the matrix $\hat M$,
which corresponds to the eigenvalue $M_1=1$, and we use normalization
$\bar {\bf e}_1(t)\cdot {\bf e}_1(t) = 1$. The matrix element
$\kappa_{11}(t,t_{\rm i})$ is periodic, $\kappa_{11}(t+\tau_F,t_{\rm i}) =
\kappa_{11}(t,t_{\rm i})$.

Equation of motion for $\kappa_{11}(t,t_{\rm i})$ for large $t-t_{\rm i}$ follows
from Eq.~(\ref{kappa}),
\begin{eqnarray}
\label{lambda_11}
{\partial\over \partial t}\left[\kappa_{11}(t,t_{\rm i}){\bf e}_1(t)\right]
&=&\mu_{11}(t)\kappa_{11}(t,t_{\rm i}){\bf e}_1(t),\nonumber\\
\mu_{11}(t)&=&
\bar{\bf e}_1(t)\cdot\hat\mu(t){\bf e}_1(t).
\end{eqnarray}

Close to the bifurcation point, the component of $\delta {\bf q}$
along the vector ${\bf e}_1(t)$ has a slowly varying factor. In
contrast, the components of $\delta {\bf q}$ along the vectors ${\bf
e}_{\nu>1}$ are ``fast''. Over time $\agt \tr^{(0)}$ they reach
quasiperiodic values for a given value of the slow component, and then
fluctuate with amplitude $\propto D^{1/2}$. From
(\ref{near_bifurcation}), the quasiperiodic values are quadratic in
the slow component and therefore small. As a consequence, the slow
motion is indeed one-dimensional,
\begin{equation}
\label{1Dansatz}
\delta{\bf q}(t) \approx Q_1(t)\kappa_{11}(t,t_{\rm i}){\bf e}_1(t).
\end{equation}
The instant $t_{\rm i}$ here is arbitrary; $Q_1(t)$ contains a
multiplicative factor that depends on $t_{\rm i}$ (but $\delta {\bf
q}(t)$ is independent of $t_{\rm i}$). The time
$t_{\rm i}$ drops out of all final expressions, see Sec.~IV.

The equation for $Q_1(t)$ is obtained by substituting
Eq.~(\ref{1Dansatz}) into Eq.~(\ref{near_bifurcation}) and then
multiplying Eq.~(\ref{near_bifurcation}) by the vector $\bar {\bf
e}_1(t)$ from the left. This gives
\begin{eqnarray}
\label{1Dequation}
&\kappa_{11}(t,t_{\rm i})\dot Q_1 = {\cal K}(Q_1,t)+ \bar{\bf e}_1(t)\cdot{\bf f}(t), \\
&{\cal K}(Q_1,t)= {1\over 2}\kappa_{11}^2(t,t_{\rm i})Q_1^2\bigl({\bf e}_1(t)\cdot
\bm\nabla\bigr)^2K_1
+\delta\!A\,(\partial_A K_1), \nonumber
\end{eqnarray}
where $K_1=\bar{\bf e}_1\cdot{\bf K}$.

In the absence of noise, the solution of Eq.~(\ref{1Dequation}) is a
sum of smooth and oscillating parts, $Q_1(t)= Q^{\rm sm}(t)+ Q^{\rm osc}$.
The term $Q^{\rm sm}$ remains nearly constant on the time scale $\tau_F$,
whereas $\dot Q^{\rm osc}\sim \omega_FQ^{\rm osc}$. It is seen from
Eq.~(\ref{1Dequation}) that $Q^{\rm osc}\propto \delta\!A$. The term
$Q^{\rm sm}$ is much larger. An equation for $Q^{\rm sm}$ can be
obtained by averaging Eq.~(\ref{1Dequation}) over time. It has the
form
\begin{equation}
\label{slow_var}
\dot{Q}^{\rm sm} = \alpha' (Q^{\rm sm})^2 + \beta'\,\delta\!A + f'(t).
\end{equation}

The coefficients $\alpha',\beta'$ in Eq.~(\ref{slow_var}) are given by
the expressions
\begin{eqnarray}
\label{coefficients}
\alpha'&=&{1\over 2}\bigl\langle \kappa_{11}(t,t_{\rm i})\bigl({\bf
e}_1(t)\cdot \bm\nabla\bigr)^2K_1\bigr\rangle_t, \nonumber\\
\beta'&=&\bigl\langle\kappa_{11}^{-1}(t,t_{\rm i})\partial K_1/\partial
A\bigr\rangle_t,
\end{eqnarray}
where $\bigl\langle\cdot\bigr\rangle_t$ means period-average
centered at time $t$,
\begin{equation}
\label{averaging}
\bigl\langle {\cal G}\bigr\rangle_t =
\tau_F^{-1}\int\nolimits_{t-\tau_F/2}^{t+\tau_F/2}dt'\; {\cal G}(t',t_{\rm i}).
\end{equation}
The result of the averaging (\ref{averaging}) is independent of $t$ for
time-periodic ${\cal G}$, as in the case of the coefficients
$\alpha',\beta'$, and therefore $\alpha',\beta'$ are independent of
$t$.

The function $f'(t)$ in Eq.~(\ref{slow_var}) is a
random force,
\begin{equation}
\label{filtered_noise}
f'(t)=\kappa_{11}^{-1}(t,t_{\rm i}) \bar{\bf e}_1(t)\cdot{\bf f}(t).
\end{equation}

Eq.~(\ref{slow_var}) has the same form as the equation for the soft
mode in the adiabatic limit (\ref{q_1_ad_}) in the absence of the term
$\propto (\omega_Ft)^2$.  For $\alpha'\beta'\,\delta\!A< 0$ the
system has a stable and an unstable stationary solution
$Q^{\rm sm}_{a,b}$. These solutions are given by an equation similar to
Eq.~(\ref{ad_states_}),
\begin{equation}
\label{nonad_states}
Q^{\rm sm}_{a,b} =\mp\sgn(\alpha')(-\beta'\,\delta\!A/\alpha')^{1/2}
\end{equation}
(in what follows without lost of generality we set $\alpha' > 0$).

Typical values of $Q^{\rm sm}$ are $\propto |\delta\!A|^{1/2}$, as seen from
Eq.~(\ref{nonad_states}). They largely exceed the amplitude of the
fast variables in $\delta {\bf q}$, which are all $\propto \delta\!A$,
in the neglect of noise. The relaxation time of $Q^{\rm sm}$ is $\tr=
|2\alpha Q^{\rm sm}_a|^{-1}\propto |\delta\!A|^{-1/2}$. It is much larger
than $\tau_F$ close to the bifurcation point. The condition
$\omega_F\tr \gg 1$ was the major approximation made in the derivation
of Eqs.~(\ref{1Dequation}), (\ref{slow_var}), besides the condition of
the weak noise.

A transformation from $Q^{\rm sm}, t$ to reduced variables
\begin{eqnarray}
\label{fast_Q}
Q=\alpha'^{1/2}Q^{\rm sm},\qquad \tau=\alpha'^{1/2}t
%
\end{eqnarray}
allows us to write Eq.~(\ref{slow_var}) in the compact form
(\ref{slow_reduced}). The random force $\tilde f(\tau)=f'(\alpha'^{1/2}t)$
is effectively $\delta$-correlated. From Eqs.~(\ref{corr_matrix}),
(\ref{filtered_noise}), its intensity is
\begin{eqnarray}
\label{noise_fast}
\tilde D = &&|\alpha'/4|^{1/2}\int_{-\infty}^{\infty}dt_1\,
\bigl\langle\kappa_{11}^{-1}(t+t_1,t_{\rm i})\kappa_{11}^{-1}(t+t_2,t_{\rm i})
\nonumber\\
&&\times \bar{\bf e}_1(t+t_1)\cdot\hat\varphi(t_1- t_2)\bar{\bf
e}_1(t+t_2)\bigr\rangle_t.
\end{eqnarray}
Here, $\hat\varphi$ is the matrix of the noise correlation functions
(\ref{corr_matrix}).

As a result of the period-averaging over $t$, in
Eq.~(\ref{noise_fast}) the integrand becomes a function of $t_1-t_2$,
and therefore the integral over $t_1$ is independent of $t_2$. Still,
it depends on $t_{\rm i}$, but this dependence will drop out of the
final expressions for observable quantities, in particular the
activation energy of escape.

\end{document}